\documentclass[preprint,12pt,authoryear]{elsarticle}

\usepackage[a4paper, top=2.5cm, bottom=2.5cm, left=1.8cm, right=1.7cm]{geometry}
\usepackage{amsmath}

\usepackage{xcolor}

\usepackage{amssymb}
\usepackage{lineno}
\usepackage{subfig}
\usepackage{float}

\journal{Elsevier}

\begin{document}

\begin{frontmatter}

\title{Enhancement of adhesion strength through microvibrations: modeling and experiments}

\author[inst1]{Michele Tricarico}
\author[inst1,inst2]{Michele Ciavarella}
\author[inst1,inst2,inst3]{Antonio Papangelo}
\affiliation[inst1]{organization={Politecnico di Bari, Department of Mechanics Mathematics and Management, TriboDynamics Lab},
            addressline={Via Orabona 4}, 
            city={Bari},
            postcode={70125}, 
            country={Italy}}

\affiliation[inst2]{organization={Hamburg University of Technology, Department of Mechanical Engineering},
            addressline={Am Schwarzenberg-Campus 1}, 
            city={Hamburg},
            postcode={21073}, 
            country={Germany}} 
\affiliation[inst3]{email: antonio.papangelo@poliba.it}

\begin{abstract}
High-frequency micrometrical vibrations have been shown to greatly influence the adhesive performance of soft interfaces, however a detailed comparison between theoretical predictions and experimental results is still missing. Here, the problem of a rigid spherical indenter, hung on a soft spring, that is unloaded from an adhesive viscoelastic vibrating substrate is considered. The experimental tests were performed by unloading a borosilicate glass lens from a soft PDMS substrate excited by high-frequency micrometrical vibrations. We show that as soon as the vibration starts, the contact area increases abruptly and during unloading it decreases following approximately the JKR classical model, but with a much increased work of adhesion {with respect to its thermodynamic value}. We find that the pull-off force increases with the amplitude of vibration up to a certain saturation level, which appeared to be frequency dependent. Under the hypothesis of short range adhesion, a lumped mechanical model was derived, which, starting from an independent characterization of the rate-dependent interfacial adhesion, predicted qualitatively and quantitatively the experimental results, without the need of any adjustable parameters.
\end{abstract}


\begin{highlights}
\item Adhesion in presence of microvibrations is studied experimentally and numerically
\item The model accurately predicts the experimental results without adjustable parameters
\item The pull-off force increases with the vibration amplitude, then saturates
\item Once the vibrations are switched on the contact area jumps abruptly 
\item {The indenter dynamics strongly influences the adhesion strength}

\end{highlights}


\end{frontmatter}


\section{Introduction} \label{sec:intro}
Traditional adhesives, such as glues or tapes, exhibit static adhesion properties---once applied, their adhesive strength remains constant or degrades over time. The design of adhesive interfaces has been inspired by nature \citep{Duan2023}, as animals like geckos \citep{gorb2007biomimetic, kamperman2010functional, li2022robust}, frogs \citep{li2024harnessing, wang2019strong}, and octopuses \citep{giordano2024mechanochromic, mazzolai2019octopus} exhibit remarkable adhesion capabilities that can be modulated according to environmental conditions \citep{arzt2021functional}. Drawing inspiration from these natural models, researchers have developed numerous artificial designs that achieve similar functionality \citep{Qin2024}. Regulating adhesion is a key requirement in advanced applications such as robotics \citep{shintake2018soft, Cacucciolo2022}, object manipulation \citep{giordano2024mechanochromic, trivedi2008soft, qu2024advanced}, human-robot interaction \citep{edsinger2007human}, crawling robots \citep{chen2020soft}, and wearable devices \citep{Huang2024}. Recent approaches to regulate adhesion in a predictable manner rely on stimuli-responsive materials, which respond to light \citep{Liu2023}, heat \citep{linghu2024fibrillar}, pH \citep{Narkar2019}, or electric \citep{Cacucciolo2022} and magnetic fields \citep{Zhao2022}.

Commonly, grippers and pads are fabricated using soft polymers, such as elastomers and silicone \citep{giordano2024mechanochromic, mazzolai2019octopus, arzt2021functional}, which behave as viscoelastic materials. Being "soft" allows the gripper to conform gently to the "rigid" counter-surface, maximizing the intimate contact area between the gripper and the object. Hence, the viscoelastic nature of soft polymers has been exploited to regulate, and generally enhance, adhesion \citep{afferrante2022effective, maghami2024viscoelastic, Maghami2024, papangelo2023detachment, violano2022size, Violano2021rate}. 

Indeed, viscoelastic materials dissipate energy when subjected to time-varying loads, which results in toughening of the interface \citep{greenwood2004theory, persson2005crack}. From a theoretical perspective, a soft viscoelastic contact can be described as an external crack propagating or healing at the interface between the indenter and the substrate \citep{schapery1975theory1, greenwood1981mechanics, greenwood2004theory, persson2005crack, ciavarella2021comparison}. According to viscoelastic crack propagation theories, due to viscoelastic losses, the effective work of adhesion $\Delta\gamma_{eff}$ depends on the crack tip speed; in particular, $\Delta\gamma_{eff}$ increases during propagation and decreases during healing, relative to the thermodynamic work of adhesion (or surface energy) $\Delta\gamma_0$. Due to this time-dependent toughening mechanism, the adhesion strength of soft interfaces depends on the loading history \citep{papangelo2023detachment,mandriota2024enhancement,afferrante2022effective,violano2022size}, unloading rate \citep{Maghami2024,Violano2021rate}, layer thickness \citep{maghami2024viscoelastic}, indenter geometry \citep{afferrante2022effective, papangelo2023detachment}.

However, for common robotic tasks such as pick-and-place manipulation, locomotion, and grasping, there is a growing need for more versatile adhesive systems that allow for the tuning of interfacial adhesion on a short time scale. For instance, a gecko can switch from high to negligible adhesion in approximately 15 ms \citep{autumn2002mechanisms}, which allows for effectively climbing and running over vertical surfaces and ceilings. 

In 2020, \citet{Shui2020} reported, for the first time, that interfacial adhesion between a soft viscoelastic substrate and a glass spherical indenter can be rapidly modulated by inducing microvibrations in the substrate at frequencies in the order of hundreds of Hertz. The basic mechanism inducing adhesion regulation relies on the dependence of the effective work of adhesion on crack speed \citep{Argatov2024, CiaTricPap2024,mandriota2024enhancement}. The externally imposed mechanical vibration causes a rapidly oscillating contact radius, which, due to the viscoelastic losses, increases the interfacial toughness. \citet{Shui2020} reported a 77-fold enhancement of the pull-off force, and a reduction to zero, by accurately tuning the vibration amplitude and frequency. In their work \citet{Shui2020} concentrated more on the dependence of the detachment force (the "pull-off" force) on the vibration amplitude $X_b$ and frequency $f$, nevertheless several aspects of the contact experiment remained  unclear, such as the evolution in time of the contact radius and contact force. Additionally, they provided a tentative model for estimating the pull-off force, resulting in a modified JKR adhesion model \citep{johnson1971surface} with an added frequency-dependent term. Nevertheless, how the estimation of the pull-off force depends on the adhesive characteristics of the soft interface remained somehow unclear from their derivation. 

More recently, \citet{Yi2024} utilized these findings to design a soft spherical-tipped adhesive finger, exploring the effect on the pull-off force of parameters like surface roughness, Young's modulus, and sphere radius. The model they propose is based on a splitting of the contact force in an adhesive and repulsive contribution, nevertheless, to fit their experimental data, they 
needed a different set of fitting parameters depending on the radius of the soft spherical indenter.  

This work aims at providing the contact mechanics scientific community with a comprehensive description of vibroadhesion, i.e. the problem of adhesion of soft contacts subjected to rapid mechanical microvibrations, by a point-by-point comparison between new experimental results and the proposed mechanical model. For this purpose, the case of a spherical indenter suspended on a compliant spring and in contact with a vibrating soft substrate is considered. In Section \ref{sec:mod}, the mechanical model describing the contact problem is introduced, which resulted in an extension of the model proposed by \citet{Shui2020}. In Section \ref{sec:num_res}, qualitative numerical results are shown to highlight the most important features of the contact problem, such as the time evolution of the contact radius and contact force during loading and unloading, the dependence of the pull-off force on the vibration amplitude and frequency, the behavior of the contact assembly at resonance. In Section \ref{sec:exp}, the experimental set-up is introduced, along with the independent characterization of the rate-dependent adhesive behavior of the soft PDMS. Then, we present a detailed comparison between the model predictions and the experimental data, which we found to be in very good agreement without the need for any adjustable parameters. Section \ref{sec:dis} discusses the experimental and  numerical results, highlighting limitations and possible future developments. Finally, closing remarks will be given in Section \ref{sec:conc}.

\begin{figure}[t]
\centering
\includegraphics[width=3.5in]{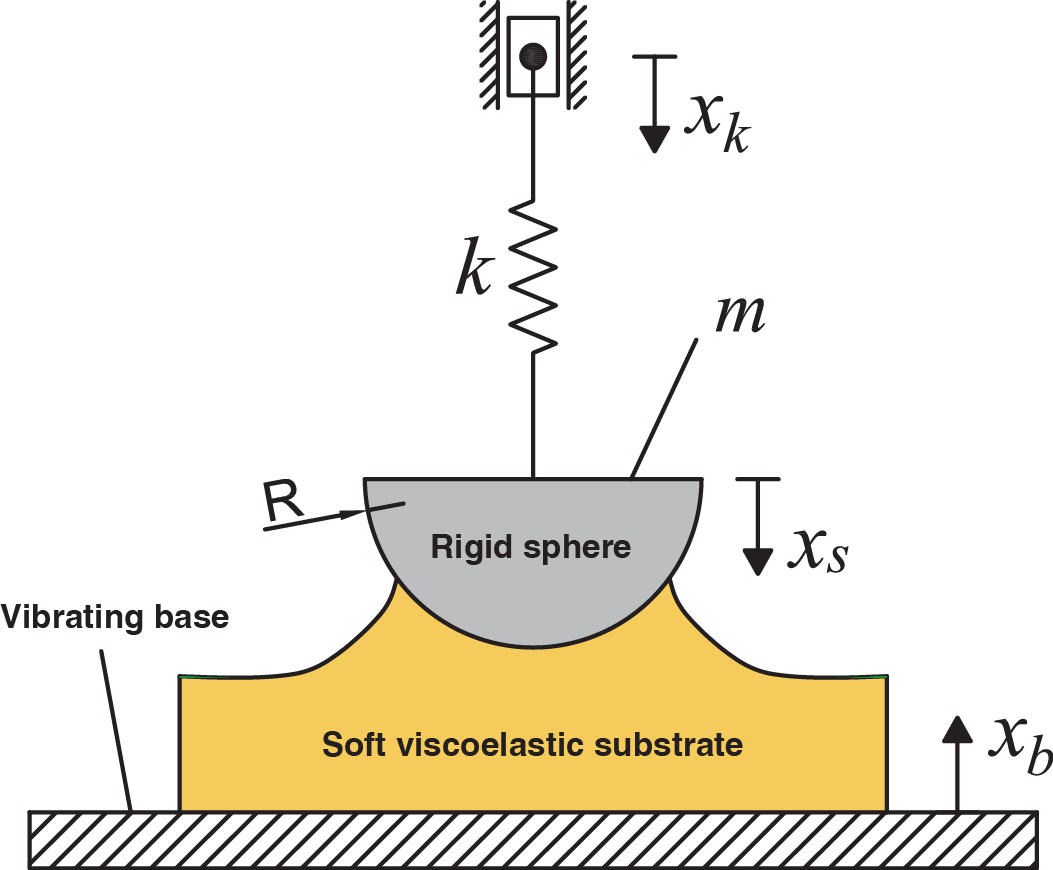}
\caption{Scheme of the modeled set-up.} \label{fig:fig1}
\end{figure}

\section{Model formulation}\label{sec:mod}

\subsection{Overview of the mechanical model}
Let us consider the problem of a rigid spherical indenter of mass $m$ and radius $R$ supported by a spring of stiffness $k$ and in contact with a soft
viscoelastic substrate. The latter is vibrating with a known (externally imposed) harmonic displacement $x_{b}\left(  t\right)  =X_{b}\sin\left(  2\pi
ft\right)  $, being $f$ the frequency, $t$ the time and $X_{b}$ the amplitude of the vibration. The position of the upper end of the spring is denoted by $x_{k}\left(  t\right)  $, while the lower end is connected to the indenter whose position is denoted by $x_{s}\left(  t\right)  $ (see Fig.
\ref{fig:fig1}). To account for the internal dissipation of the assembly a damping
coefficient $c$ is also introduced in the equilibrium equation, so that it reads%
\begin{equation}
m\overset{\cdot\cdot}{x}_{s}+c\overset{\cdot}{x}_{s}+k(x_{s}-x_{k})=F_{c}+mg
\end{equation}
where a dot superposed means differentiation with respect to time (e.g.
$\overset{\cdot}{x}_{s}=dx_{s}/dt$), $F_{c}$ is the contact force that the substrate
applies to the indenter, considered positive when adhesive, and $g=9.807$
m/s$^{2}$ is the gravitational acceleration. For the contact model, short range
adhesion is assumed, which is appropriate for modeling the contact with soft
elastomers \citep{tabor1977surface}. In this framework, the JKR theory \citep{johnson1971surface} applies, which gives the following expressions for the contact force $F_{c}$ and indentation depth $\delta$:
\begin{align}
F_{c}  &  =-\frac{4E^{\ast}a^{3}}{3R}+\sqrt{8\pi\Delta\gamma_{0}E^{\ast}a^{3}%
}\label{JKRforce}\\
\delta &  =\frac{a^{2}}{R}-\sqrt{\frac{2\pi\Delta\gamma_{0}a}{E^{\ast}}}
\label{JKRdelta}%
\end{align}
where, $E^{\ast}=\left(  \left(  1-\nu_{1}^{2}\right)  /E_{1}+\left(
1-\nu_{2}^{2}\right)  /E_{2}\right)  ^{-1}$ is the contact modulus, being $\left\{  E_{i},\nu_{i}\right\}  _{i=1,2}$ the Young's moduli and the Poisson's ratios of the two bodies in contact at the given excitation frequency $f$, $a$ is the contact radius and $\Delta\gamma_{0}$ is the thermodynamic work of adhesion.

It is assumed that the indenter reference position corresponds to the case of static equilibrium when subjected to its own weight and to the adhesive force corresponding to zero indentation ($\delta=0$), hence for $x_{k}=0$ one has%
\begin{equation}
x_{s0}=\frac{F_{c,\delta=0}+mg}{k}%
\end{equation}
By the change of variable $x=x_{s}-x_{s0}$ one can rewrite the equilibrium equation as%

\begin{equation}
m\overset{\cdot\cdot}{x}+c\overset{\cdot}{x}+k(x-x_{k})=F_{c}\left(
\delta\right)  -F_{c,\delta=0} \label{Equi}%
\end{equation}

where $F_{c,\delta=0}=\left(  4/3\right)  \pi R\Delta\gamma_{0}$. The normal contact force $F_{c}\left(  \delta\right)  $ is an implicit equation of the indentation $\delta$ through the JKR model Eq.s (\ref{JKRforce},\ref{JKRdelta}).

It is assumed here that the substrate vibration, physically applied at the
bottom surface of the viscoelastic substrate (see Fig. \ref{fig:fig1}), is rigidly transferred to the whole substrate, implying that also the reference free surface of the soft substrate moves as $x_{b}\simeq X_{b}\sin\left(  2 \pi f t\right)  $ (this implies we are neglecting inertia forces within the substrate), hence the indentation $\delta\left(  t\right)  $ is written as
\begin{equation}
\delta\left(  t\right)  =x\left(  t\right)  +X_{b}\sin\left(  2\pi ft\right)
\label{delta}%
\end{equation}
Notice that only when the indenter displacement is negligible (i.e., $x\left(
t\right)  /x_{b}\left(  t\right)  \approx0$) the sphere indentation will be approximately equal to the substrate oscillation, i.e. $\delta\left(
t\right)  \approx x_{b}\left(  t\right)  $.\\

\subsection{Rate-dependent adhesion}
Elastomers and silicones behave as viscoelastic materials, which requires careful modeling to account for the dissipative contribution of the bulk \citep{persson2005crack,Creton2016,schapery1975theory1,schapery1975theory2,schapery2022theory,greenwood1981mechanics,greenwood2004theory}, which may tremendously affect the tackiness of soft materials, as it has been shown for several geometries, including the case of a flat punch \citep{maghami2024viscoelastic,papangelo2023detachment} and of a spherical indenter \citep{Maghami2024,vandonselaar2023silicone,muser2022crack,violano2022size,tiwari2017effect}.

Nevertheless, if we limit our interest to soft elastomers and short range
adhesion, dissipation may be assumed to be concentrated close to the crack tip
(along the contact periphery), and be accounted for by considering an
effective work of adhesion $\Delta\gamma_{eff}$. The latter represents an enhancement of the
thermodynamic work of adhesion $\Delta\gamma_{0}$ when the crack
propagates, and a reduction when the crack heals \citep{persson2005crack,greenwood1981mechanics,greenwood2004theory,nazari2024friction,Papangelo2024friction, carbone2022theory}. Several theories have tried to relate the internal dissipation mechanisms of the viscoelastic material to the
effective adhesive energy \citep{persson2005crack,greenwood1981mechanics,carbone2022theory}. However, the comparison with the
experimental data has remained satisfactory only in a limited range of crack speeds, perhaps because of other nonlinear rate-dependent mechanisms acting within the process zone \citep{Maghami2024}, only recently considered in theoretical models \citep{barthel2024linear,persson2024influencetemperaturecracktipspeed}. Nonetheless, there exists a strong experimental evidence that
when the crack is propagating, the effective surface energy increases as a
function of the crack speed as a power law \citep{barthel2024linear, hui2022steady}, in accordance with the well known phenomenological model firstly introduced by
\citet{gent1972effect}\footnote{Due to the time-temperature superposition in viscoelastic behavior of soft materials the reference velocity can be written as $v_0=(\kappa a_T^n)^{-1}$ where $\{\kappa,n\}$ are constants with $0<n<1$ and $a_T$ is the WLF factor that shifts the viscoelastic modulus at different temperatures \citep{WLF}.}:%

\begin{equation}
\Delta\gamma_{eff}=\left\{
\begin{array}
[c]{cc}%
\Delta\gamma_{0}\left(  1+\left\vert \frac{v}{v_{0}}\right\vert ^{\alpha
}\right)   & \qquad v\geq0, \text{opening}\\
\Delta\gamma_{0}\left(  1+\left\vert \frac{v}{v_{0}}\right\vert ^{\alpha
}\right)  ^{-1} & \qquad v<0, \text{closing}
\end{array}
\right.  \label{GS}%
\end{equation}
where $v=-da/dt$ is the crack speed, $\left\{  v_{0},\alpha\right\}  $ are positive characteristic constants of the interface. We have considered that
the effective work of adhesion is reduced for closing (healing) cracks $\left(
v<0\right)  $ with $\left[  \Delta\gamma_{eff}%
/\Delta\gamma_{0}\right]  _{\text{closing}}\approx\left[  \Delta\gamma
_{eff}/\Delta\gamma_{0}\right]  _{\text{opening}}^{-1}$, which is generally a satisfactory approximation, as shown by \citep{greenwood2004theory}. 

\subsection{Dimensionless notation}
Following \citep{maugis2013contact},\ let us introduce the following reference contact radius $a_{r}$ and indentation $\delta_{r}$%

\begin{equation}
a_{r}=\left(  \frac{3\pi\Delta\gamma_{0}R^{2}}{4E^{\ast}}\right)
^{1/3};\text{\qquad}\delta_{r}=\frac{a_{r}^{2}}{R}=\left(  \frac{9\pi
^{2}\Delta\gamma_{0}^{2}R}{16E^{\ast2}}\right)  ^{1/3};
\end{equation}
and consequently the dimensionless quantities%

\begin{gather}
\widetilde{a}=\frac{a}{a_{r}},\text{\quad\quad}\widetilde{\delta}=\frac
{\delta}{\delta_{r}},\text{\quad\quad}\widetilde{x}=\frac{x}{\delta_{r}%
},\text{\quad\quad}\widetilde{x}_{s/b/k}=\frac{x_{s/b/k}}{\delta_{r}%
},\text{\quad\quad}\widetilde{X}_{b}=\frac{X_{b}}{\delta_{r}},\\
\widetilde{F}_{c}=\frac{F_{c}}{\pi R\Delta\gamma_{0}},\text{\quad\quad}\widetilde{t}=\omega_{n}t,\text{\quad\quad}\widetilde
{k}=\frac{k}{\frac{4}{3}E^{\ast}a_{r}},\text{\quad\quad}\widetilde
{f}=\frac{2\pi f}{\omega_{n}},\text{\quad\quad}\widetilde{\Delta\gamma}%
_{eff}=\frac{\Delta\gamma_{eff}}{\Delta\gamma_{0}},
\end{gather}
where $\omega_{n}=\sqrt{{k}/{m}}$ is the natural frequency of the linear oscillator when it is not in contact. Hence, the JKR contact model equations, for the effective work of adhesion, become%

\begin{align}
\widetilde{F}_{c}  &  =-\widetilde{a}^{3}+\sqrt{6\widetilde{a}^{3}%
\widetilde{\Delta\gamma}_{eff}}\label{JKRdlessFc}\\
\widetilde{\delta}  &  =\widetilde{a}^{2}-\frac{2}{3}\sqrt{6\widetilde
{a}\widetilde{\Delta\gamma}_{eff}} \label{JKRdlessdelta}%
\end{align}
and the equilibrium equation is written as%

\begin{equation}
\widetilde{x}^{\prime\prime}+2\zeta\widetilde{x}^{\prime}+(\widetilde
{x}-\widetilde{x}_{k})=\frac{1}{\widetilde{k}}\left(  \widetilde{F}_{c}\left(
\widetilde{\delta}\right)  -\widetilde{F}_{c,\widetilde{\delta}=0}\right)
\label{equidless}%
\end{equation}
where $\zeta={c}/({2\sqrt{km}})$ is the dimensionless damping ratio, the prime represents differentiation with respect to the dimensionless
time, i.e. $\widetilde{x}^{\prime}=d\widetilde{x}/d\widetilde{t}$. Hence, substituting
$\widetilde{\delta}=\widetilde{x}+\widetilde{X}_{b}\sin\left(  \widetilde
{f}\widetilde{t}\right)  $ into the JKR contact model (Eq.
(\ref{JKRdlessdelta})) gives the dimensionless effective surface energy%

\begin{equation}
\widetilde{\Delta\gamma}_{eff}=\frac{3}{8\widetilde{a}}\left(  \widetilde
{a}^{2}-\widetilde{x}-\widetilde{X}_{b}\sin\left(  \widetilde{f}\widetilde
{t}\right)  \right)  ^{2} \label{dgammaeff}%
\end{equation}
and substituting into Eq. (\ref{JKRdlessFc}) the contact force becomes%

\begin{equation}
\widetilde{F}_{c}=-\widetilde{a}^{3}+\frac{3}{2}\widetilde{a}\left(
\widetilde{a}^{2}-\widetilde{x}-\widetilde{X}_{b}\sin\left(  \widetilde
{f}\widetilde{t}\right)  \right)  \label{Fcdless}%
\end{equation}
Using Eq. (\ref{Fcdless}) into the equilibrium equation Eq. (\ref{equidless})
and rearranging the Gent and Shultz law (Eq. (\ref{GS})), the system dimensionless evolution equations can be written as:%

\begin{equation}
\left\{
\begin{array}
[c]{l}%
\widetilde{x}^{\prime\prime}+2\zeta\widetilde{x}^{\prime}+(\widetilde
{x}-\widetilde{r}\widetilde{t})=\frac{1}{\widetilde{k}}\left[  -\widetilde
{a}^{3}+\frac{3}{2}\widetilde{a}\left(  \widetilde{a}^{2}-\widetilde
{x}-\widetilde{X}_{b}\sin\left(  \widetilde{f}\widetilde{t}\right)  \right)
-\frac{4}{3}\right] \\
\widetilde{a}^{\prime}=\left\{
\begin{array}
[c]{lc}%
-\widetilde{v}_{0}\left(  \frac{3}{8\widetilde{a}}\left(  \widetilde{a}%
^{2}-\widetilde{x}-\widetilde{X}_{b}\sin\left(  \widetilde{f}\widetilde
{t}\right)  \right)  ^{2}-1\right)  ^{1/\alpha} & \qquad\text{, if }%
\widetilde{a}^{\prime}<0\\
\widetilde{v}_{0}\left(  \frac{8\widetilde{a}}{3}\left(  \widetilde{a}%
^{2}-\widetilde{x}-\widetilde{X}_{b}\sin\left(  \widetilde{f}\widetilde
{t}\right)  \right)  ^{-2}-1\right)  ^{1/\alpha} & \qquad\text{, if
}\widetilde{a}^{\prime}>0
\end{array}
\right.
\end{array}
\right.  \label{SystEq}%
\end{equation}
where $\widetilde{a}^{\prime}=-\widetilde{v}$.

It was assumed that, starting from a certain initial position, the spherical indenter is unloaded at a constant rate $r$, hence $\widetilde{x}_{k}=\widetilde
{r}\widetilde{t}$, being $\widetilde{r}=r/\left(  \omega_{n}\delta_{r}\right)
$, negative during unloading. 

The system of ODEs in Eq. (\ref{SystEq}) was solved in MATLAB$^{\copyright }$ using the ode function \textit{ode23t}, which can handle moderately stiff dynamical problems and it provides the integration
results without the use of numerical damping. For a given preload
$\widetilde{F}_{c0}$ the initial contact radius $\widetilde{a}_{0}$ and indenter position $\widetilde{x}_{0}$ are found, while it was always assumed that the unloading phase started when the
indenter was at rest, hence $\widetilde{x}_{0}^{\prime}=0$. {The numerical simulations were terminated when, during unloading, the contact radius fell below 0.}

\section{Model predictions}
\label{sec:num_res}

Before comparing the model predictions against the experimental data, we give here an overview of the system adhesive behavior to varying amplitude and frequency of the imposed microvibrations. In this section the results will be presented in dimensionless form. Otherwise differently stated, it was assumed $\widetilde{k}=0.0774$, $\widetilde{F}_{c0}=-6.547$, $\widetilde{r}=-0.1$ and for the Gent and Schultz parameters $\widetilde{v}_{0}=9.168\times10^{-4}$ and $\alpha=0.531$.

\subsection{Influence of frequency and amplitude}
\begin{figure}[H]
\centering
\includegraphics[width=7in]{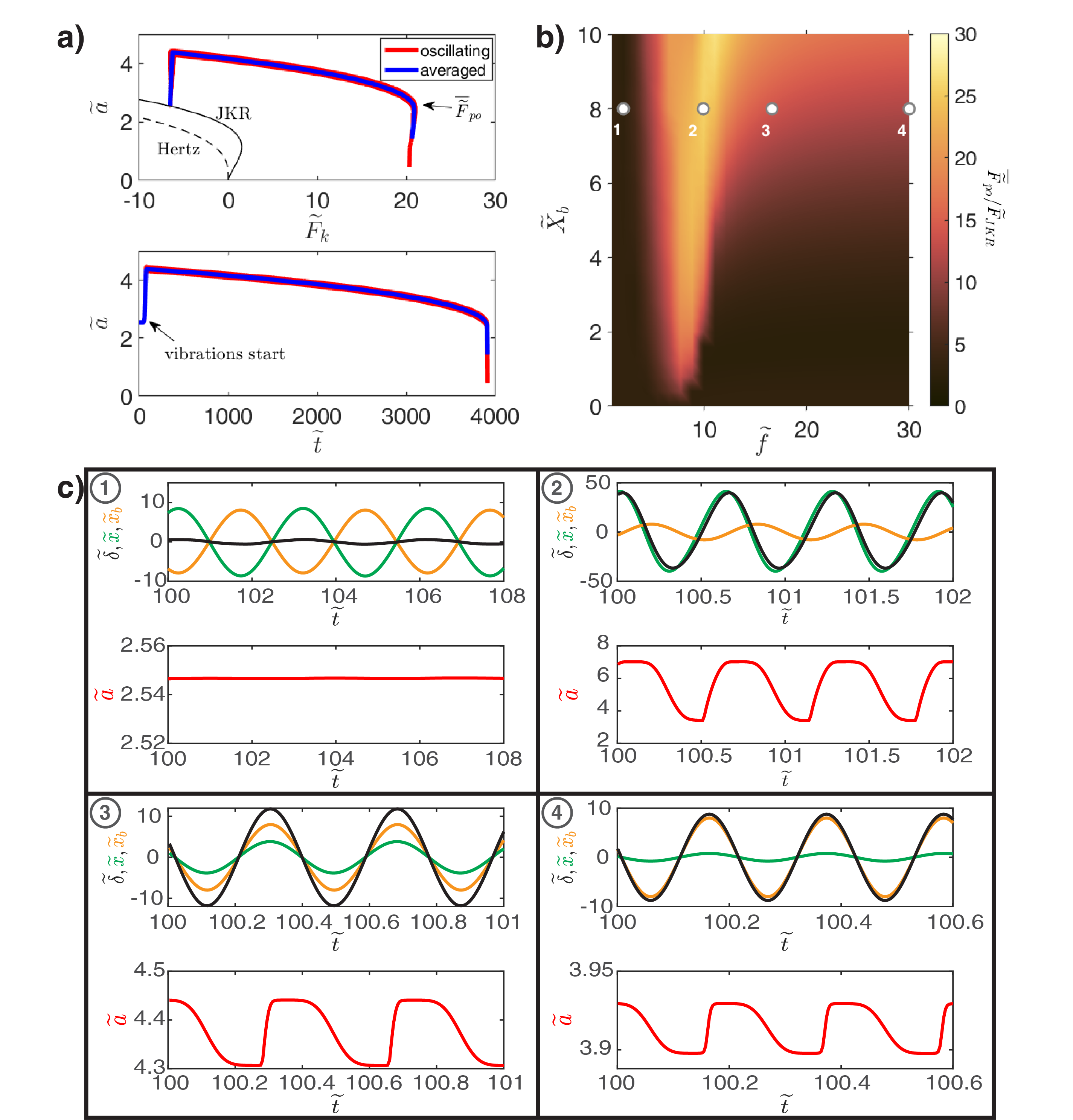}
\caption{a) Top panel: contact radius $\widetilde{a}$ versus the spring reaction force
$\widetilde{F}_{k}$ in dimensionless form. The black solid line is the JKR
curve, the black dashed line is the Hertz contact model, the red solid line is the oscillating $\widetilde{a}$ versus $\widetilde{F}_{k}$ curve, the blue
solid line is its average over one vibration period denoted as $\overline
{\widetilde{a}}$ vs $\overline{\widetilde{F}}_{k}$. 
Bottom panel: contact radius $\widetilde{a}$ versus the dimensionless
time $\widetilde{t}$ curve (the red line is the oscillating curve, the blue
line is its average value over one vibration period). The curves were obtained by fixing $\widetilde{X}_{b}=8$ and $\widetilde{f}=16.62$. b) Map of the pull-off force $\widetilde{F}_{po}$ normalized with respect to the JKR quasi-static value $\widetilde{F}_{JKR}=3/2$ as a function of the
dimensionless vibration amplitude $\widetilde{X}_{b}$ and frequency
$\widetilde{f}$. c)
Time evolution of the sphere indentation $\widetilde{\delta}$ (black lines) position $\widetilde{x}$ (green lines), substrate vibration $\widetilde{x}_{b}$ (orange lines) and contact radius $\widetilde{a}$ (red lines). The four panels (numbered from 1 to 4) correspond to the locations highlighted in the map in b). The vibration amplitude is $\widetilde
{X}_{b}=8$ for every case, while the excitation frequency is: (1)\ $\widetilde{f}=2.1$,
(2)\ $\widetilde{f}=9.9$, (3) $\widetilde{f}=16.6$, (4) $\widetilde
{f}=30.0$.} \label{fig:fig2}
\end{figure}

To qualitatively show the system behavior, let us simulate a dynamic adhesion test. The indenter is firstly slowly driven into contact with the substrate, up to a preload of $\widetilde{F}_{c0}=-6.547$ (negative in compression), which is then kept constant for a certain dwell time to allow for full relaxation of the viscoelastic substrate. At this point the substrate starts to vibrate and, after a certain dwell time, the indenter is unloaded at a constant unloading rate $\widetilde{r}$. Fig. \ref{fig:fig2}a shows the evolution of the dimensionless contact radius $\widetilde{a}$ versus the spring reaction $\widetilde{F}%
_{k}=\widetilde{k}(\widetilde{x}-\widetilde{x}_{k})+4/3$ (defined as positive when tensile).
The dimensionless vibration amplitude and frequency were set to $\widetilde
{X}_{b}=8$ and $\widetilde{f}=16.62$, respectively. 

As we are considering high frequency microvibrations, whose time-scale is much smaller than the time-scale relative to the indenter unloading, it is useful to define averaged (over one vibration period $\widetilde{T}=2\pi/\widetilde{f}$) values of the contact force ($\overline
{\widetilde{F}}_{c}$) and contact radius ($\overline{\widetilde{a}}$).
{Considering the indenter is slowly unloaded, $\overline
{\widetilde{F}}_{c}$ coincides with the averaged spring reaction force $\overline{\widetilde{F}}_{k}$:}

\begin{equation}
\frac{1}{\widetilde{T}}\int_{0}^{\widetilde{T}}\widetilde{F}_{c}\left(
\widetilde{t}\right)  d\widetilde{t}\approx\frac{1}{\widetilde{T}}\int
_{0}^{\widetilde{T}}\widetilde{F}_{k}\left(  \widetilde{t}\right)
d\widetilde{t}%
\end{equation}

Figure \ref{fig:fig2}a (top panel) shows in red the curve $\widetilde{a}$
vs $\widetilde{F}_{k}$, in blue the averaged value $\overline{\widetilde{a}}$  vs $\overline{\widetilde{F}}_{k}$, while the solid and dashed black lines
refer to the JKR and Hertz contact model, respectively. As soon as the
substrate starts vibrating the contact radius jumps to a larger value. This jump is clearly visible in Fig. \ref{fig:fig2}a (bottom panel) where the evolution of the dimensionless contact radius as a function of
the dimensionless time is shown. Notice that $\widetilde{a}\left(
\widetilde{t}=0\right)$ coincides with the quasi static JKR value at the given preload $\widetilde{F}_{c0}$. The base vibration starts at
$\widetilde{t}=50$, while the unloading  phase is set to start at $\widetilde{t}=100$. The detachment point is considered where $\overline{\widetilde{F}}_{k}$ is maximum, which was defined as $\overline{\widetilde{F}}_{po}=\max\left(  \overline{\widetilde{F}}%
_{k}\right)  $.  In the example considered, the pull-off force is increased by a factor $\simeq15$ with respect to the quasi-static counterpart $(\widetilde{F}_{JKR})$. 

Having clarified the typical loading-unloading scenario, the effect of $ \widetilde{X}_{b}$ and $\widetilde{f}$ on the pull-off force enhancement $\overline{\widetilde{F}}_{po}/\widetilde{F}_{JKR}$ is shown in the map in Fig. \ref{fig:fig2}b. {In general the pull-off force $\overline{\widetilde{F}}_{po}$ increases when the vibration amplitude is increased, while the behavior with respect to the frequency is influenced by the nonlinear stiffness introduced by the contact $k_c$. At a given indentation $\overline{\delta}_0$ the contact stiffness is $k_c=\left(d F_{c}/d\delta \right)\vert _{\delta=\overline{\delta}_0}$, which depends on the indentation $\overline{\delta}_0$ as the force versus indentation curve is nonlinear. The corresponding dimensionless contact stiffness will be $\widetilde{k}_c=k_c/(\frac{4}{3}E^*a_r)$. In the numerical simulations the contact model is always satisfied, i.e. there is no contact loss except at the pull-off point where the simulation is truncated. Clearly too large vibration amplitudes may cause detachment of the indenter as soon as the vibrations start. This regime was not investigated here as it may lead to cycles of jump-in and -out of contact including impacts, which we were not interested in.}

Indeed, substituting in the equation of motion (Eq. (\ref{Equi})) the expressions for $\delta$ (Eq. (\ref{delta})) and its derivatives, in the limit of very high
frequency and with low apparatus stiffness and damping, which is our case of interest
($\widetilde{k}\ll1$ and $\zeta\ll1$), one obtains the resonance frequency of
the system at a given mean indentation $\overline{\delta}_0$ to be $\omega
_{nc}=\sqrt{\left(  \left.  \frac{dF_{c}\left(  \delta\right)  }{d\delta
}\right\vert _{\delta=\overline{\delta}_0}\right)  /m}$ (see
Appendix B).
For a typical contact radius of $\widetilde{a}=5.1$ one finds $\widetilde{k}_c=7.1$ and $\widetilde
{f}_{nc}=\omega_{nc}/\omega_{n}\simeq9.6$ that is consistent with the
numerical results found. 

Notice that in the JKR model, the relation contact force versus indentation is an implicit equation of the contact radius $\widetilde{a}$,
nevertheless an approximation is to consider the stiffness of a flat punch of
radius $\widetilde{a}$ that, in our case, would estimate the contact stiffness
as $\widetilde{k}_{c}=\left(  3/2\right)  \widetilde{a}\simeq7.6$, giving $\widetilde
{f}_{nc}\simeq10$ within the
$5\%$ error. {Scrutiny of the map in Fig. \ref{fig:fig2}b would also reveal the stiffening behavior of the contact as $k_c$ increases with the indentation hence the locus of the maxima of the pull-off force bends towards
the right due to the increase of the system resonance frequency $\omega_{nc}$.}

A better understanding of the system dynamical behavior at different frequencies can be obtained by checking the time evolution of the following variables (Fig. \ref{fig:fig2}c): sphere indentation $\widetilde{\delta}$ (black lines) and position $\widetilde{x}$ (green lines),
substrate vibration $\widetilde{x}_{b}$ (blue lines) and contact radius $\widetilde{a}$ (red lines). Four cases were analysed, each corresponding to a representative value of frequency (highlighted in Fig. \ref{fig:fig2}b): case 1)\ $\widetilde{f}=2.1$,
case 2)\ $\widetilde{f}=9.9$, case 3) $\widetilde{f}=16.6$, case 4) $\widetilde{f}=30.0$. The vibrations amplitude was fixed to $\widetilde{X}_{b}=8$ for every case.
Notice that, for the sake of an easier comparison, only the oscillating part of the signals $\widetilde{\delta}\left(  \widetilde{t}\right)  $ and $\widetilde{x}\left(
\widetilde{t}\right)  $ is shown, i.e. their mean values was subtracted.

Prior to resonance (case 1), indentation and base displacement move in-phase (consider the convention of signs in Fig. \ref{fig:fig1} meaning that as the substrate moves the indenter follows. This results in a minimal variation of indentation depth $\widetilde{\delta}$ and in a very small variation of the contact radius $\widetilde{a}$, preventing
adhesion enhancement. Case 2 describes the system behavior near the resonance: the indenter vibration this time dominates with respect to the base
excitation. This implies $\widetilde{\delta}\approx\widetilde{x}$, which
are almost in quadrature of phase with respect to the base excitation $\widetilde{x}_b$. Large, high frequency, oscillations of $\widetilde{\delta}$ result in large
oscillations of the contact radius, yielding a significant amplification of the
adhesive force. 
By increasing the excitation frequency above the resonance frequency {($\widetilde{f}_{nc}\approx10$)}, the indentation and base displacement move out-of-phase (consider
the convention of signs in Fig. \ref{fig:fig1}), which signifies that the base
vibration is amplified by the oscillator motion resulting in $\widetilde
{\delta}>\widetilde{x}_{b}$ (cases (3, 4)). Hence, within our
modeling assumptions, we have $\lim_{\widetilde{f}\rightarrow+\infty}\widetilde
{\delta}(\widetilde{f})=\widetilde{x}_{b}$, resulting in a very mild dependence on the excitation frequency, as it can be clearly seen in Fig. \ref{fig:fig2}b.%

\begin{figure}[t]
\centering
\includegraphics[width=\textwidth]{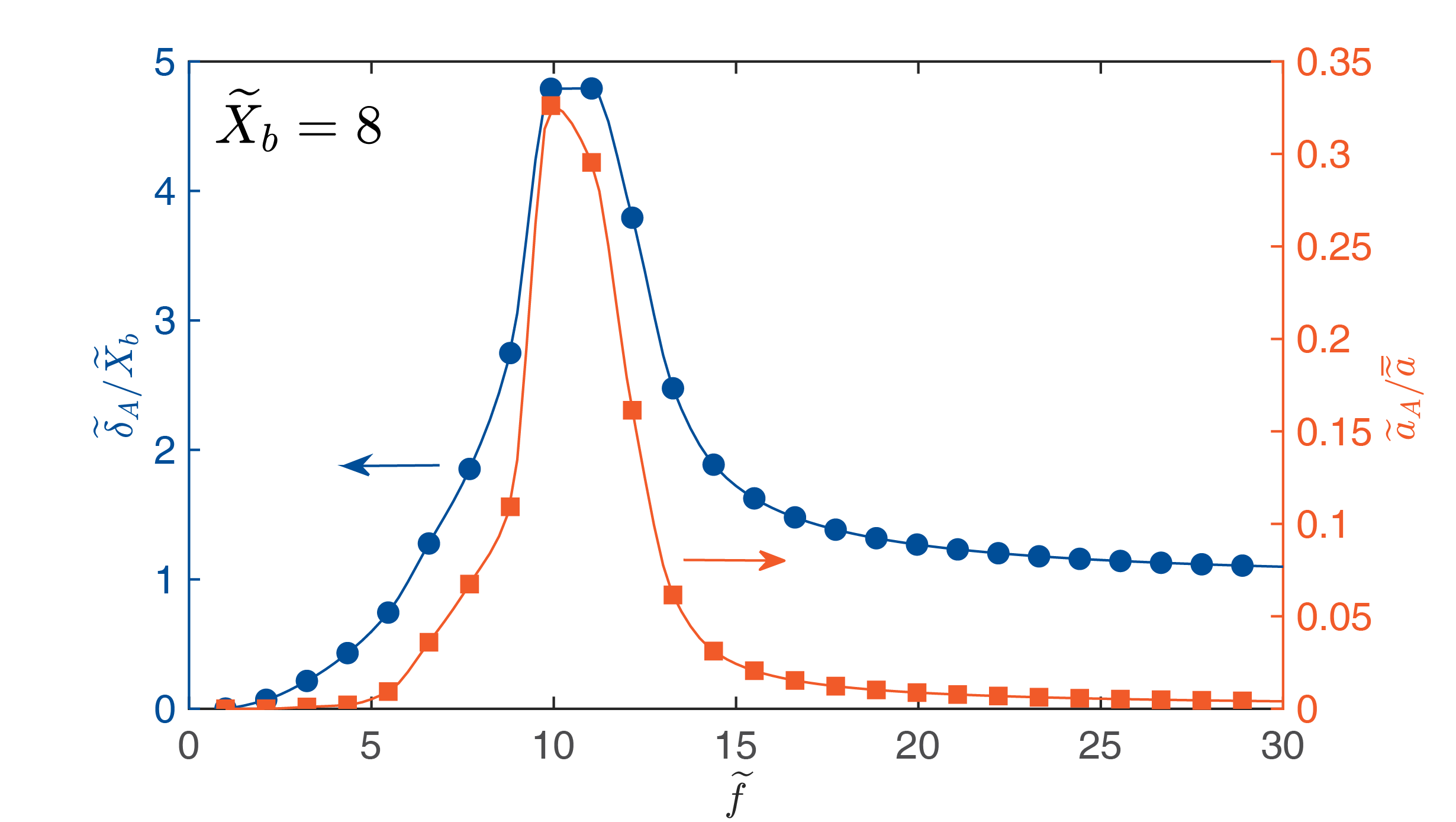}
\caption{ Normalized vibration amplitude for the contact radius ($\widetilde{a}%
_{A}/\overline{\widetilde{a}}$, left y-axis, blue curve) and indentation
($\widetilde{\delta}_{A}/\widetilde{X}_{b}$, right y-axis, orange curve) as a
function of the frequency $\widetilde{f}$ for $\widetilde{X}_{b}=8$. The
quantities are estimated when the vibrations are switched on and the sphere is subjected to the preload $\widetilde{F}_{c0}=-6.547$.} \label{fig:fig3}
\end{figure}

\subsection{Behaviour at pull-off}
To clarify the mechanism behind the adhesion enhancement, Fig.
\ref{fig:fig3} shows how the oscillating parts of the contact radius and
indentation (taken as soon as the vibrations are switched on) vary with the excitation frequency, for a fixed amplitude $\widetilde{X}_{b}=8$. The representative quantities  $\widetilde{a}_{A}=\left(  \max\left(
\widetilde{a}\right)  -\min\left(  \widetilde{a}\right)  \right)  /2$ and
$\widetilde{\delta}_{A}=\left(  \max\left(  \widetilde{\delta}\right)
-\min\left(  \widetilde{\delta}\right)  \right)  /2$, normalized by the mean contact radius $\overline{\widetilde{a}}$ and the substrate vibration amplitude $\widetilde{X}_{b}$, respectively, are plotted as functions of the frequency. Both curves show the same qualitative behavior, with
the oscillating part of the contact radius and indentation increasing
at low frequencies, then reaching a peak at the resonance frequency and later
decreasing as the frequency is further increased. Notice that the contact area
variation is predicted to be quite severe at resonance, being $\widetilde
{a}_A\approx0.33\ast\overline{\widetilde{a}}$, while this oscillation steeply
drops moving away from the resonance frequency and remaining at high frequency
within the $1\%$ of the average contact radius. This is consistent with the
experimental observation in \citet{Shui2020}. Qualitatively
similar conclusions can be drawn for the variation of $\widetilde{\delta}_{A}$. Notice that as the frequency increases, $\widetilde{\delta}%
_{A}/\widetilde{X}_{b}\rightarrow1$. At very high frequency $\widetilde{f}\gg\widetilde{f}_{nc}$, a simplified model for the system in Fig. \ref{fig:fig1} can be considered that neglects the indenter dynamics, which is briefly presented in \ref{sec:appendixC}.  

\section{Experimental tests} \label{sec:exp}

\subsection{Description of the experimental test-rig}

\begin{figure}[t]
\centering
\includegraphics[width=6in]{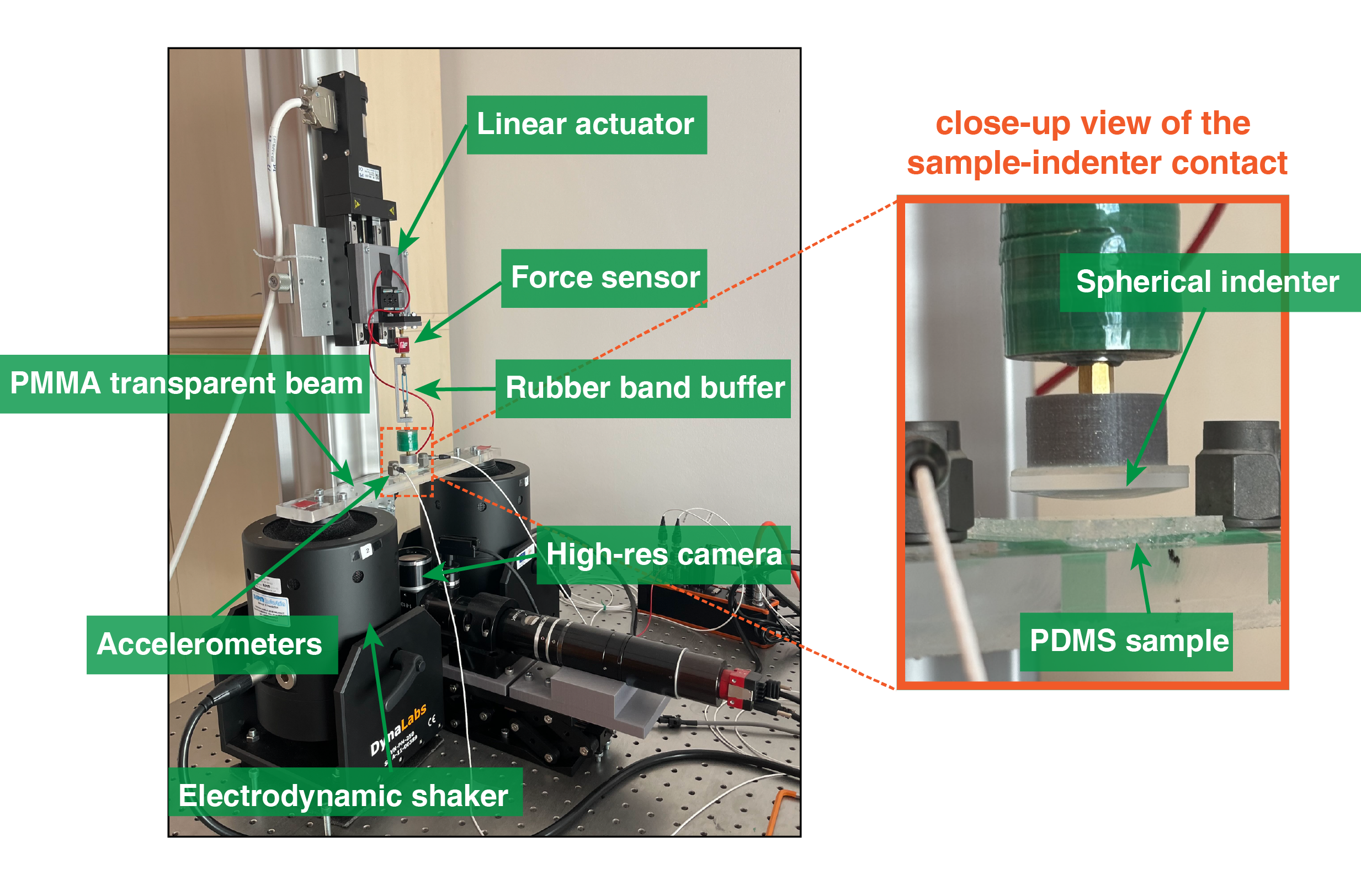}
\caption{Experimental test-rig used for the dynamic tests.} \label{fig:fig4}
\end{figure}

The customized test rig shown in Fig. \ref{fig:fig4} was used to conduct the vibration-regulated adhesion experiments. The indenter, a borosilicate crown glass plano-convex spherical lens with radius of curvature $R=51.5$ mm, was connected to a uniaxial S-beam load cell (Futek LSB205, natural frequency 1180 Hz, resolution 1 mN). In order to shield the load cell from highly oscillating axial loads during the testing, a compliant rubber band buffer was positioned between the load cell and the indenter. Additional weights were placed on top of the indenter to keep the rubber band in tension. {The stretched rubber band was $4$ cm long, with a squared cross section of $1 \times 1$ mm$^2$. It was connected on both ends (one end to the load cell, the other to the indenter) through steel eye-bolts. The calibration was made by fixing the end connected to the indenter and pulling on the rubber band by moving the upper end connected to the load cell. Force-displacement curves were recorded using the same loading rate of the vibration tests, in a range of forces similar to the one employed in the tests, where the force-displacement curve is approximately linear.  A stiffness of $k=320$ N/mm was estimated.} \\
A micrometrical motorized linear stage (Physik Instrumente, M-403.2DG) was used to drive the assembly in the direction normal to the PDMS surface.

The dynamic excitation was supplied by two electrodynamic shakers (Dynalabs, DYN-PM-250), that also served as supports for the PMMA transparent beam where the PDMS sample (fixed onto a glass slide) was placed. The sample was positioned right in the center of the beam and two piezoelectric accelerometers (DYTRAN 3055D1, frequency range 1-10000 Hz) where mounted at its sides, allowing to measure the acceleration of the PMMA beam close to the PDMS substrate. The vibration amplitude was then measured by double-integration of the acceleration signal. The measured signals were acquired and elaborated through a data acquisition system (Dewesoft, DAQ system KRYPTON-6XSTG). A high-resolution camera (Alvium 1800 U-2040m, resolution 20 Mpixel, pixel size 2.74 $\mu$m, frame rate 10 fps, {shutter exposure time $5$ ms}), captured live images of the contact patch through a right angle optical element positioned beneath the contact. {The purpose of the imaging was to measure the averaged values of the contact radius during the whole test. The contact edge was sharp enough to clearly distinguish an average contact circle in most of the tests. Some blur was visible only at large amplitudes or near the pull-off.}

By using a custom code built in Wolfram Mathematica\textsuperscript{\textcopyright} to post-process the images, the corresponding evolution of the contact radius was determined. 

The standard loading protocol was the following: the sample was loaded with a $F_{c0}=-250$ mN preload (loading rate $100$ $\mu$m/s) and a dwell time of $60$ s was allowed for material relaxation. Subsequently, the substrate vibration was switched on by starting the electrodynamics shakers and the target amplitude was reached by manually adjusting the amplifier’s gain. An additional $60$ s dwell time was considered in order to stabilize all the acquired signals, e.g. force, vibration amplitude, contact radius. The sample was finally unloaded at $r=-5$ $\mu$m/s. All the experimental tests were performed at room temperature $\approx20^\circ C$.

\begin{figure}[t]
\centering
\includegraphics[width=5.5in]{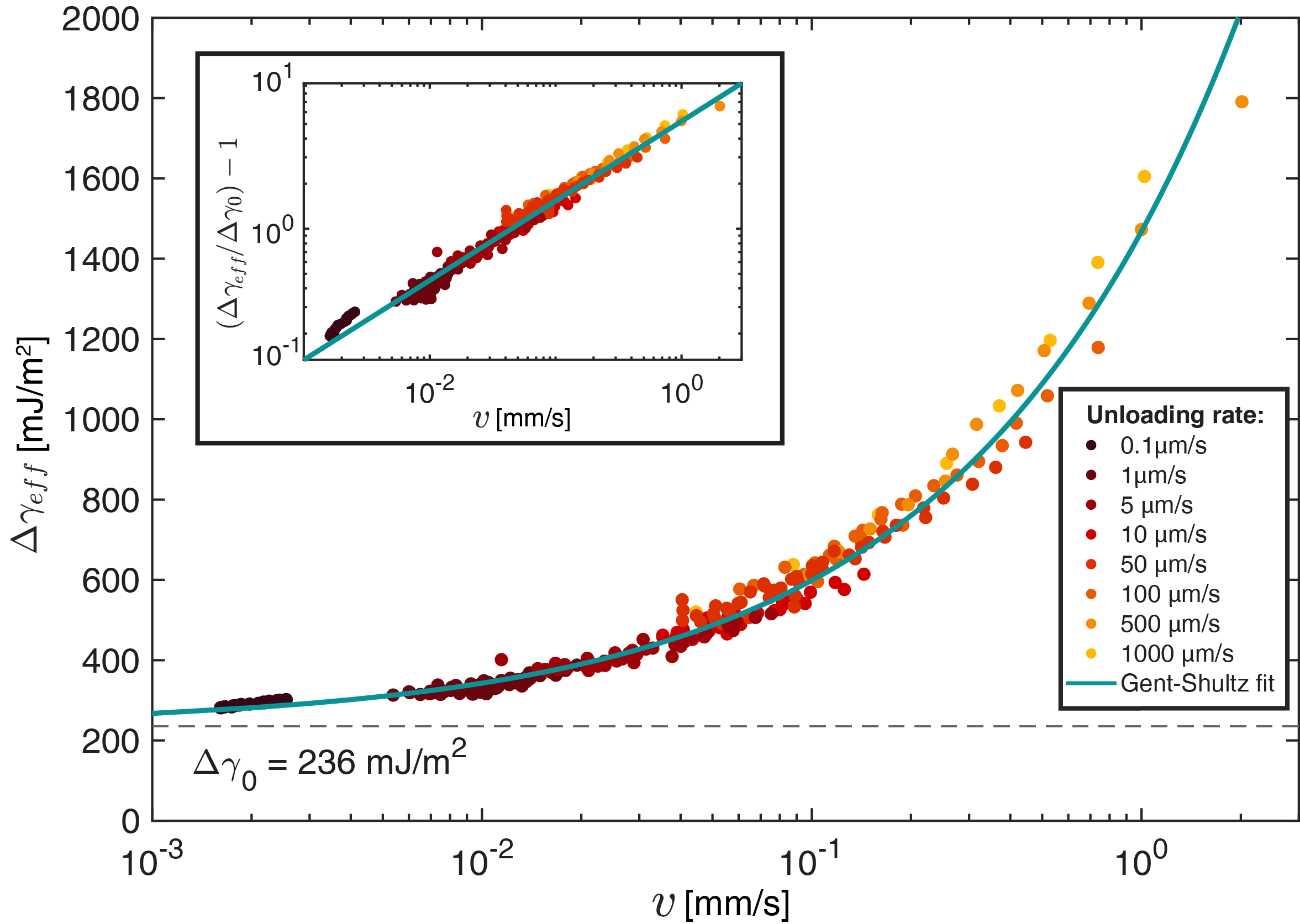}
\caption{Characterization of the interface rate-dependent adhesion. The effective work of adhesion $\Delta \gamma_{eff}$ as a function of the crack speed obtained from the experimental data (markers) through Eq. (\ref{GSexp}). The data points were fitted using Eq. (\ref{GS}) (solid line), resulting in a reference speed of $v_0=0.0445$ mm/s and an exponent $\alpha=0.531$. The inset shows a log-log representation of the quantity $(\Delta \gamma_{eff}/\Delta \gamma_{0}-1)$ that highlights the power law scaling.} \label{fig:fig5}
\end{figure}

\subsection{PDMS sample preparation}
The PDMS samples utilized in the adhesion experiments were fabricated using the commercial elastomer Dow Sylgard 184®. The base:curing agent volume ratio was set to 10:1. The mixture was degassed under vacuum in a desiccator for 15 minutes at room temperature. The resin was subsequently poured into a petri dish containing a glass slide, previously activated by air plasma to favor intimate bonding without the use of any glue or intermediate layer. Finally, the resin was cured for an initial 18 h period at room temperature, followed by a second 90 min period at 75°C. {Such a two-step protocol was adapted from \cite{delplanque2022solving} and allows to reduce the shrinkage-induced deformation of the PMDS slab upon curing, preserving a flat surface.} The final  dimensions of the PDMS sample were 2.5 x 2.5 cm in plane and 3 mm thickness.

\subsection{Independent characterization of the interface rate-dependent adhesion}

Before to move to the dynamic experiments, the rate-dependent adhesive behavior of the PDMS-glass interface was performed. Firstly, the quasi-static properties ($E_{0}^{\ast}$, $\Delta\gamma_{0}$) were determined by a classical JKR indentation test. The indenter was loaded at $r=0.1$ $\mu$m/s, then the contact was stabilized for a dwell time of 60 s and later the unloading phase started with $r=-0.1$ $\mu$m/s (refer to \ref{sec:appendixA} for further details). Further reduction of the loading/unloading rate below $|r|<0.1$ $\mu$m/s did not provide any significant difference in the results obtained. By fitting the quasi-static unloading curve contact radius versus contact force with a JKR model (Figure \ref{fig:figAppA1}), we found $\Delta \gamma_0 = 236$  $\text{mJ/m}^2$ and a relaxed contact modulus $E_{0}^{\ast}=2.8$ MPa. {For the same material \cite{oliver2023adhesion} found $( E^*_0=1.49$ MPa, $\Delta\gamma_0 = 141$ mJ/m$^2$) and \cite{dorogin2018contact} found $( E^*_0=2.2$ MPa, $\Delta\gamma_0 = 280$ mJ/m$^2$). We believe the slightly larger value of Young moduls compared to \cite{oliver2023adhesion} and \cite{dorogin2018contact} may be due to the finite thickness of the PDMS substrate.}
The rate-dependent adhesion behavior of the interface was characterized by post-processing the results of several adhesion tests conducted at unloading rates in the range $r\in \left[0.1,...,1000 \right]$ $\mu$m/s, where unloading started from a fully relaxed substrate. Writing the JKR contact model (Eq. (\ref{JKRforce})) with an effective work of adhesion $\Delta \gamma_{eff}$, one finds

\begin{equation}
\Delta \gamma_{eff}  = \frac{(F_H-F_c)^2}{6 \pi R F_H} \label{GSexp}
\end{equation}

where $F_H= {4 E_0^\ast a^3}/(3R)$ is the Hertzian load, $F_c$ is the normal contact force as measured by the load cell and $a$ is the measured contact radius. By fitting the experimental data with the empirical Gent-Shultz power law model (Eq. \ref{GS}) we found the reference speed to be $v_0=0.0445 \;$ mm/s and $\alpha=0.531$, as shown in Fig. \ref{fig:fig5}. The inset in Fig. \ref{fig:fig5} shows the quantity $(\Delta \gamma_{eff}/\Delta \gamma_{0}-1)$ in log-log scale to highlight the power law scaling of the effective surface energy with respect to the crack speed. 

In summary, the dimensional parameters of the system are reported in Table \ref{tab:tab1}. Notice that the natural frequency $\omega_n$ and the damping ratio $\zeta$ of the free oscillator (not in contact) were obtained performing a free vibration test of the indenter. In Table \ref{tab:tab1}, the effective contact modulus $E^*_{eff}=4.5$ MPa is reported that is a factor $\simeq1.6$ larger than the quasi-static value determined by the JKR contact test (see \ref{sec:appendixA}). This accounts for the fact that the substrate gets stiffer when excited in the frequency range $f \in \left[200,400\right]$ Hz and agrees well with our vibrating experiments (see \ref{sec:appendixB}) and with independent DMA characterization of the PDMS material in \citet{Maghami2024}. 

\begin{table}[t]
    \centering
    \begin{tabular}{ccc}
     \textbf{System parameters}& \textbf{Symbol} & \textbf{Value} \\
        Indenter radius of curvature & $R$ & 51.5 mm \\
        Natural frequency of the free system & $\omega_n$ & 70.4 rad/s \\
        Damping coefficient & $\zeta$ & 0.0317 \\
        Rubber band stiffness & $k$ & 320 N/m\\
        Effective contact modulus of the PDMS sample& $E_{eff}^*$& 4.5 MPa\\
        Initial work of adhesion & $\Delta \gamma_0$ & 236 mJ/m$^2$ \\
        Gent-Shultz reference speed & $v_0$ & 44.5 $\mu$m/s \\
        Gent-Shultz exponent & $\alpha$ & 0.531\\
        Preload & $F_{c0}$ & -250 mN \\
        Unloading rate & $r$ & -5 $\mu$m/s \\
        Reference contact radius& $a_r$ & $6.895\times10^{-4} \;$ m\\
        Reference indentation depth& $\delta_r$ & $9.231\times10^{-6} \;$ m\\
        Vibrations amplitudes & $X_{b}$ & [20, ..., 160] $\mu$m \\
        Vibrations frequency & $f$ & [200, 300, 400] Hz \\

    \end{tabular}
    \caption{Dimensional parameters of the mechanical system.}
    \label{tab:tab1}
\end{table}

\subsection{Dynamic adhesion tests}

\begin{figure}[t]
\centering
\includegraphics[width=5in]{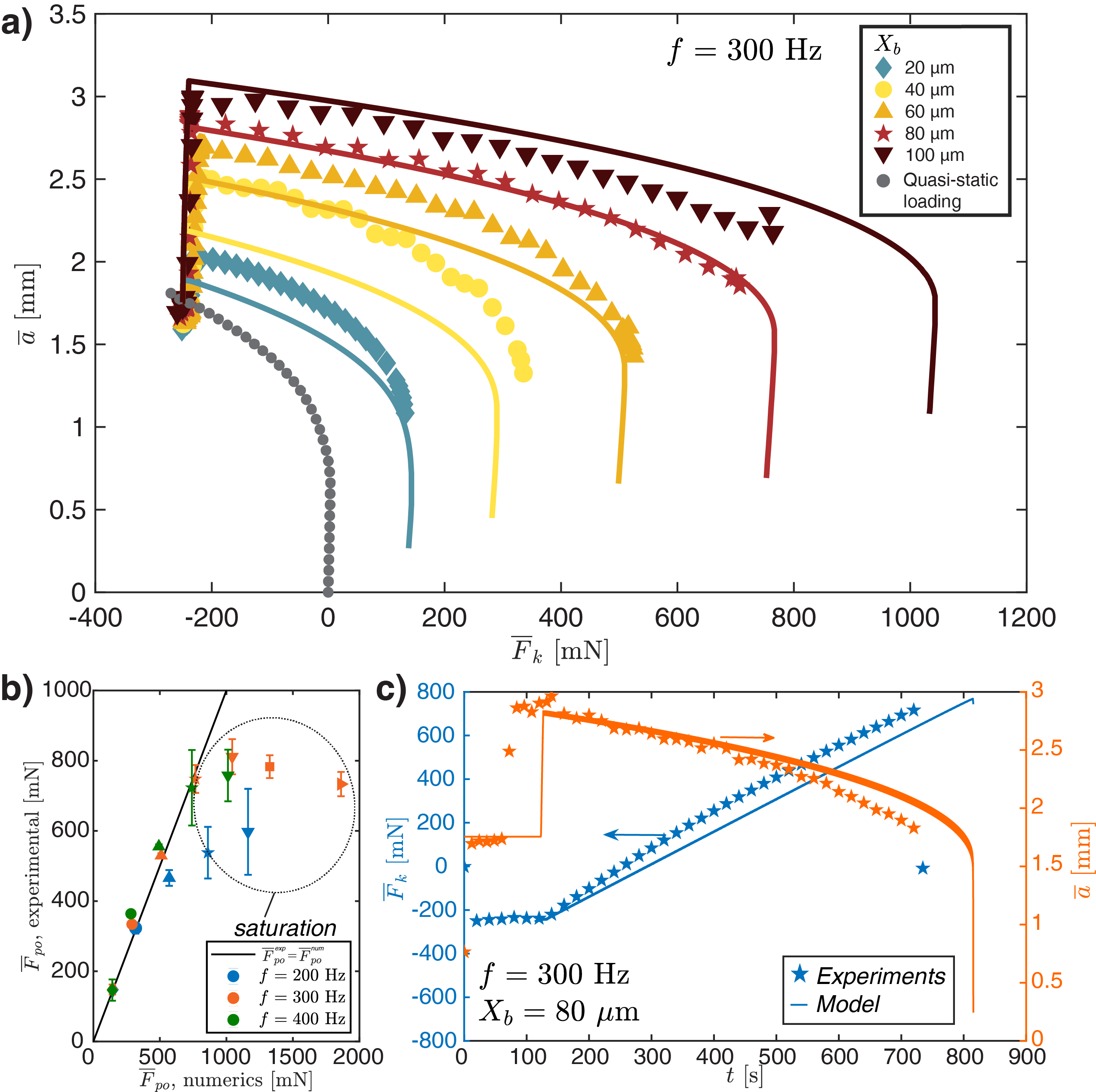}
\caption{Comparison with experiments a) Average spring reaction force ($\overline{F}_{k}$) vs. average contact radius ($\overline{a}$) curves for the dynamic tests performed with $f=300$ Hz and $X_b=[20,40,60,80,100]$ $\mu$m. b) Experimental pull-off force (3 test were performed per each value of amplitude $X_b$: data points correspond to the mean, the error bars indicate the standard deviation) vs. numerical prediction for different values of frequency ($f=[200, 300, 400]$ Hz). {The black straight line indicates the locus where the experimental pull-off force  $\overline{F}_{po}^{exp}$ is equal to the numerical pull-off force $\overline{F}_{po}^{num}$.} c) Force ($\overline{F}_{k}$) vs. time (t) and contact radius (a) vs. time (t) curves for the case ($X_b=80$ $\mu$m, $f=300$ Hz). Experiments and simulations  are matched at the time instant when unloading starts ($t\simeq120$ s).
Note that each marker shape always corresponds to the same value of $X_b$, in all the panels. The extra orange square and right-oriented triangle in panel b) refer to $X_b=120$ $\mu$m and $X_b=160$ $\mu$m, respectively} \label{fig:fig6}
\end{figure}

Having elucidated the adhesive response of the indenter without microvibrations, we move here to the dynamic tests. In what follows, the experimental results will be compared with the predictions of the model we have presented in Section \ref{sec:mod}. Starting from the physical dimensional parameters listed in Table \ref{tab:tab1}, the corresponding dimensionless parameters were derived, which are listed in Table \ref{tab:tab2}. In Figure \ref{fig:fig6}a, we report the measured average contact radius $\overline{a}$ versus the average spring reaction force $\overline{F}_k$ curves for typical dynamic tests performed at $f =300$ Hz and $X_b$ ranging from 20 to 100 $\mu$m (respectively, diamonds, circles, up-triangle, star, down-triangle). The experimental quasi-static loading curve (with no vibration) is shown as a gray line. The model predictions obtained integrating Eq. (\ref{SystEq}) are shown as solid lines, using the same color of the experimental data they refer to (Figure \ref{fig:fig6}a). 

A satisfactorily agreement is found between the model predictions and the experimental data without any adjustable parameter. We find that as soon as the microvibrations are switched-on, the contact radius increases suddenly jumps to much larger values. For example, in the case of $X_b=100$ $\mu$m, a two-fold increase, with respect to the the quasi-static value under the same normal load ($F_{c0}=-250$ mN), is observed. At the frequency of our tests, the larger the vibration amplitude the larger the pull-off force reached. Both the model and the experiments suggest that at low vibration amplitude, the detachment seems to be fostered by an instability, similarly to the mechanism happening in a classical JKR adhesion experiment. Nevertheless, when the vibration amplitude increases (see $X_b=100$ $\mu$m) we find very good agreement between the model and the experimental curve except at pull-off. Indeed, during the experiment, the detachment suddenly happened around 800 mN, while the model would predict an instability around 1100 mN. 

{In the classical JKR model, in load control, detachment happens through an instability at the turning point of the contact radius versus contact force curve where $dF_c/da$ vanishes. The experimental data at low vibration amplitude suggest the same mechanism as it can be seen from Fig. \ref{fig:fig6}a. Increasing the vibration amplitude $X_b$ the detachment in the experiments appears more abrupt, suggesting a different mechanism.}

The effect of the excitation frequency on the pull-off force was investigated in Fig. \ref{fig:fig6}b, where the pull-off data are reported as estimated from the numerical model (x-axis) and as obtained experimentally (y-axis) for the frequency $f=[200, 300, 400]$ Hz and for the vibration amplitude in the range from 20 $\mu$m to 160 $\mu$m (the same markers of Fig. \ref{fig:fig6}a are employed, with the extra square and right-oriented triangle referring to $X_b=120$ $\mu$m and $X_b=160$ $\mu$m, respectively). The effect of frequency seems negligible for small amplitudes ($X_b<80 \mu m$), but it becomes significant at large amplitudes ($X_b>80 \mu m$), with the pull-off force increasing with $f$. For small amplitudes the model matches very-well the experimental data, successfully reproducing the pull-off enhancement with increasing $X_b$. However, for larger amplitudes, experiments show a "saturation" of the pull-off that the model fails to predict (see the Discussion in Section \ref{sec:dis}).

\begin{table}[t]
    \centering
    \begin{tabular}{ccc}
     \textbf{Dimensionless parameter}& \textbf{Symbol} & \textbf{Value} \\
        Rubber band stiffness & $\widetilde{k}$ & 0.0774\\
        Gent-Shultz power law fit parameter& $\widetilde{v}_0$ & $9.168\times10^{-4}$ \\
        Preload & $\widetilde{F}_{c0}$ & -6.547 \\
        Unloading rate & $\widetilde{r}$ & -0.008 \\
        Vibrations amplitudes & $\widetilde{X}_{b}$ & $2.167 \times [1, ..., 8]$ \\
        Vibrations frequency & $\widetilde{f}$ & $8.925 \times [2,3,4]$ \\
    \end{tabular}
    \caption{Dimensionless parameters.}
    \label{tab:tab2}
\end{table}

By way of example, let us analyze the case of $f=300$ Hz and $X_b=80$ $\mu$m. Fig. \ref{fig:fig6}c shows the evolution of the average reaction force and of the average contact radius versus the time, for both the model predictions (solid lines) and the experimental data (star markers). Experiments and simulations  are matched at the time instant when unloading starts ($t\simeq120$ s). Notice that the model precisely reproduced the contact radius jump when the vibrations start, but also its time evolution and the variation of the average spring reaction force during all the unloading phase.

\begin{figure}[H]
\centering
\includegraphics[width=6in]{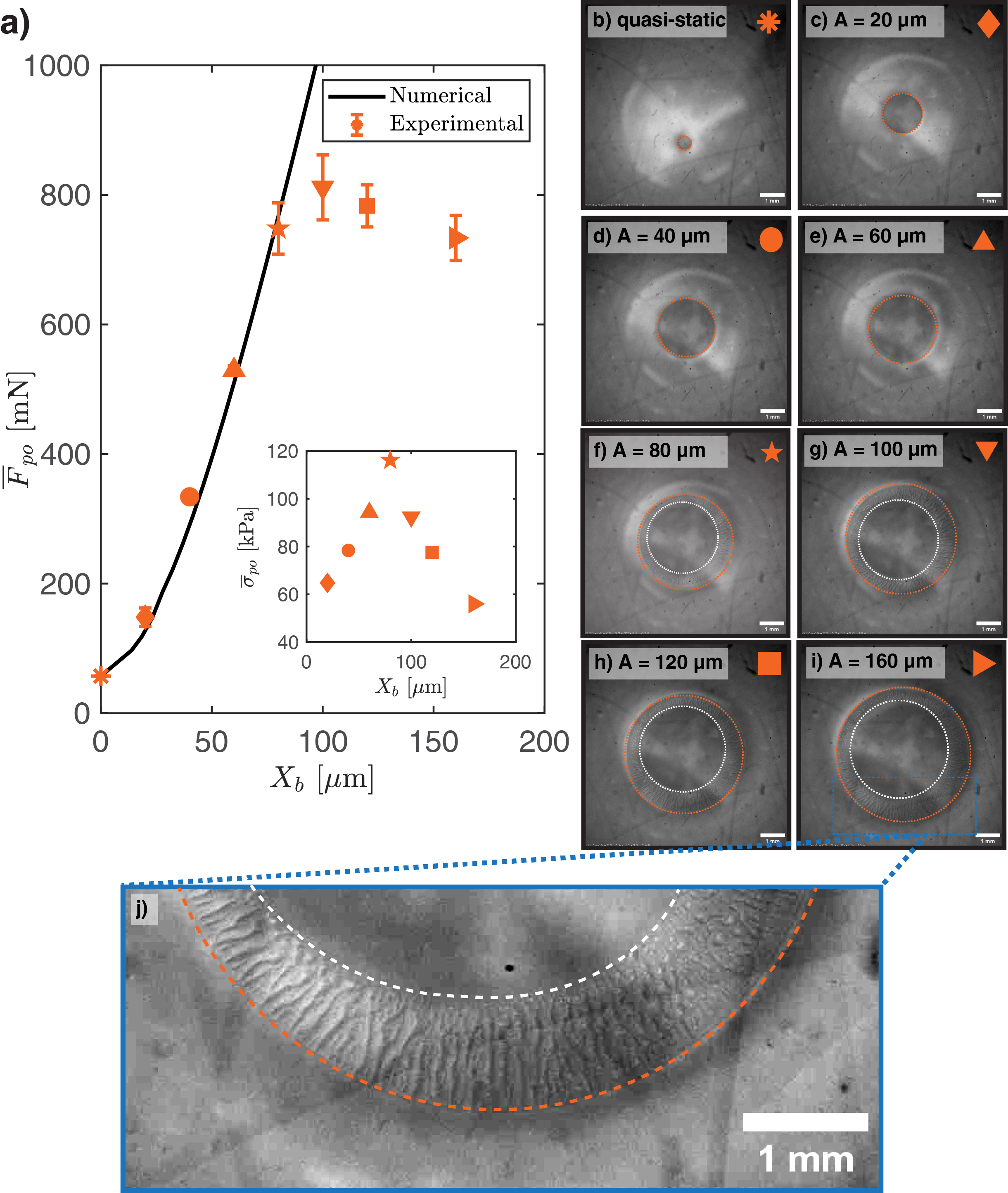}
\caption{a) Pull-off force as a function of the vibration amplitude $X_b$ (each marker corresponds to a certain value of $X_b$, consistently with Fig. \ref{fig:fig6}). The model prediction (solid line) is compared against the experimental data points. b-i) Micrographs of the contact area just prior to pull-off for each test: b) quasi-static; c)  $X_b=20$ $\mu$m; d) $X_b=40$ $\mu$m; e)$X_b=60$ $\mu$m; f) $X_b=80$ $\mu$m; g) $X_b=100$ $\mu$m; h) $X_b=120$ $\mu$m; i) $X_b=160$ $\mu$m; j) close-up view of the radial wrinkles leading to abrupt pull-off ($X_b=160$ $\mu$m). The frequency was set to 300 Hz. The contrast was enhanced to allow for a better visualization. {In panels (b-j) the the orange dashed circles delimitate the contact periphery, while in panels (f-j)
the white dashed circles indicate what we considered the intimate contact circle, i.e. not including the radial wrinkles.}} \label{fig:fig7}
\end{figure}

The behavior at pull-off was further scrutinized in Fig. \ref{fig:fig7}a, where the pull-off force is shown as a function of the vibration amplitude for $f=300$ Hz. The black solid line refers to the model predictions while the markers (each corresponding to a certain value of $X_b$, consistently with Fig. \ref{fig:fig6}) refer to the experimental data (the error-bar indicates one standard deviation over three independent measurements). Figure \ref{fig:fig7} b-i show micrographs of the contact area immediately prior to detachment, with the contact periphery indicated by a dashed orange circle. Pull-off is preceded by a significant reduction of contact radius in the quasi-static case and in the small amplitude cases. As $X_b$ increases, the detachment takes place in a more "abrupt" manner. The latter case, referring to $X_b=[100, 120, 160]$ $\mu$m, seems related to the saturation of the pull-off force that the model is unable to predict. It appears that above a certain vibration amplitude, $X_b\simeq80$ $\mu$m for the case considered in Fig. \ref{fig:fig7}a, the condition from detachment shifts from an instability point to a different mechanism. Further scrutiny of the micrographs at high vibration amplitudes reveals the presence of radial wrinkles (see close-up view for the case $X_b=160$ $\mu$m in Fig. \ref{fig:fig7}j) close to the contact periphery.
A possible interpretation is that these wrinkles are the result of local elastic instabilities on the surface due to the hoop compressive stress $\sigma_\theta$ generated during the unloading phase, which for $\nu=0.5$, are equal to the Hertzian component of the pressure distribution acting normal to the substrate interface ($\sigma_\theta = -p_0\sqrt{1-(r/a)^2}$, being $p_0$ the maximum stress in the Hertzian solution, $r$ the radial coordinate, see Eq. (4.130) in \cite{maugis2013contact}). Nevertheless, local instabilities, resulting in wrinkles and cavitation phenomena, have been studied theoretically and numerically in spherical adhesive contact. \cite{he2022modeling} observed them for quasi-static loading of very confined elastic systems where $a/h\gtrsim4$, being $h$ the thickness of the layer (while in our case we have $a/h\simeq1)$. However, our experimental set-up involves a viscoelastic material with base excitation, therefore definitive conclusions on the origin of the observed wrinkles cannot be drawn at this point.

By considering both the pull-off force and the contact area (for $X_b > 80$ $\mu$m, we considered only the patch within the white dashed circle to be in intimate contact), we estimated the average stress at pull-off, $\overline{\sigma}_{po}$, which is plotted as a function of the vibration amplitude in the inset of Fig. \ref{fig:fig7}a. The results show that $\overline{\sigma}_{po}$ increases with $X_b$, peaks at $X_b = 80$ $\mu$m, and decreases as $X_b$ increases further. This has two implications: firstly, there is an "optimal" vibration amplitude that maximizes the adhesive capabilities of the interface; {secondly, the pull-off force saturation cannot be ascribed to the limited adhesive strength of the interface, as otherwise we would have observed a plateau of $\overline{\sigma}_{po}$ at high vibration amplitudes} (a "strength-limited" mechanism, see \cite{papangelo2023detachment}). This, together with the observation of radial wrinkles, raises the question of what is the mechanism that triggers the saturation of the pull-off force at large amplitudes. This matter goes beyond the scope of the present work, and will be subject of further investigations.\\
{Furthermore, the radial asymmetry of the wrinkles (indicated by the non-concentricity of the orange and white dashed circles in Fig. 7f-j) might suggest the presence of shear stresses. \cite{oliver2023adhesion} showed that shear forces can reduce the pull-off force, nevertheless in our set-up the indenter is hung through a rubber band to the load cell like a pendulum, hence shear forces could not be equilibrated. Furthermore, shear forces introduce asymmetry into the contact area that appears more elliptical than circular \citep{sahli2019shear,papangelo2019shear}, something that we never observed in our experiments. Although a small misalignment in the set-up cannot be excluded, we believe the saturation of the pull-off force cannot be explained by the presence of shearing loads.} 

\section{Discussion}\label{sec:dis}

The problem of viscoelastic amplification of the adhesive strength of soft interfaces has been extensively discussed in the literature \citep{persson2005crack,Creton2016,schapery1975theory1,schapery1975theory2,schapery2022theory,greenwood1981mechanics,greenwood2004theory,maghami2024viscoelastic,papangelo2023detachment,Maghami2024,vandonselaar2023silicone,muser2022crack,violano2022size,tiwari2017effect, mandriota2024adhesive}. Theoretically, the maximum amplification should be of the order of the ratio between the glassy modulus $E_{\infty}$ and the rubbery modulus $E_{0}$ \citep{persson2005crack}, which for PDMS is greater than $\simeq2000$ \citep{Maghami2024}. However, in spherical probe contact experiments, amplification factors are only a fraction of the theoretical $E_{\infty}/E_{0}$, generally due to difficulties in achieving very high interfacial crack speeds. Vibroadhesion seems able to push the amplification boundaries further \citep{Shui2020}. Indeed, regardless of the generally slow unloading process, high-frequency oscillations of the probe indentation imply high-frequency oscillations of the crack tip, resulting in interfacial adhesion strengthening due to viscoelastic dissipation.  

\cite{wahl2006oscillating} and \cite{greenwood2006oscillatory} already studied oscillating adhesive viscoelastic contacts for a PDMS substrate, using a micrometric diamond tip with a radius of $10$ $\mu$m, oscillation amplitude of the normal load in the range $\sim [15, 150]$ nN, and frequencies in the range $[1
, 200]$ Hz \citep{wahl2006oscillating}. In their experiments, they mostly focused on the unexpected shift of the contact stiffness from the  JKR solution (which can be even negative) to the flat punch solution due to contact line pinning. However, they experimental results did not show any enhancement of the pull-off force, even when the excitation frequency was changed from 2 Hz to 200 Hz. We have shown here that adhesion enhancement is governed by the interfacial crack speed, and in particular in a harmonic oscillation of the contact radius we would have the maximum crack speed to be $\max(v)=2\pi f{a}_A$, being $a_A$ the amplitude of the contact radius oscillation. From our Fig. \ref{fig:fig3} we have $a_A\approx10^{-2}\overline{a}$ and typically for Hertzian contacts $\overline{a}\approx10^{-1}R$, hence $\max(v)\approx 2\pi f (10^{-3}R)$. Using, $f=100$ Hz and $R=10$ $\mu$m \citep{wahl2006oscillating} one gets $\max(v)\simeq6$ $\mu$m/s that is much smaller than the Gent and Schultz reference speed we measured ($\max(v)\approx 0.135v_0$), which may explain why they did not find adhesion enhancement. As a comparison, in our experiments we have $f=300$ Hz and $R=51.5$ mm, hence $\max(v)=2\pi f (10^{-3}R)\approx10^5$ $\mu$m/s $\gg v_0$. The possibility to enhance macroscopic adhesion through microvibrations will depend on the term $\propto(fR)$, hence smaller indenter radius will require larger excitation frequencies to achieve the same adhesion enhancement. 

Only recently, \cite{Shui2020} showed that high-frequency microvibrations could lead to significant adhesion enhancement, up to $77$ times larger than the quasi-static case. In their work, \cite{Shui2020} proposed a model to interpret their data, considering the effect of base vibration as equivalent to the application of a harmonic inertia force on the indenter. This represents the force-controlled limit of the model we have proposed here (see Appendix B), and after some approximations, they derived a force versus contact radius curve similar to the JKR model, but with an additional frequency-dependent term. From this, \cite{Shui2020} estimated the pull-off force as dependent on vibration amplitude, frequency, and damping ratio (see Eq. (\ref{ShuiFpo})). However, from a physical standpoint, the role of the damping coefficient and of the Gent and Schultz parameters on the pull-off force estimation remains somewhat unclear from their derivation. Notice that, in terms of dimensional pull-off forces, in our experiments we observed very similar values as in \cite{Shui2020} ($\max(\overline{F}_{po})\simeq811.5$ mN), nevertheless we report an enhancement of about $\overline{F}_{po}/\overline{F}_{JKR}\simeq811.5/57.3\simeq14.16$, where we selected as a baseline pull-off force the one obtained during unloading in a classical JKR adhesion quasi-static (no vibrations) test ($\Delta \gamma_{0} =236$  mJ/m$^2$, $\overline{F}_{JKR}=57.3$ mN, see Fig. \ref{fig:figAppA1}). The 77-folds enhancement reported by \cite{Shui2020}, stems from the much smaller value of the baseline surface energy that they employed, namely $\Delta \gamma_{0} =40$ mJ/m$^2$, which was measured via a peeling test. Unfortunately, full details of their quasi-static characterization of the interface were not provided.   

Very recently, a different setup for vibroadhesion experiments was developed by \cite{Yi2024}, where, unlike the setup presented here, a vibrating PDMS spherical-tipped probe with diameters $D = [1,2,3,4,5]$ mm was unloaded from a fixed rigid substrate. \cite{Yi2024} observed both amplification and reduction of the adhesive force, depending on the amplitude and frequency of the excitation. They proposed an approximate model for the pull-off force based on a splitting of the contact force into its adhesive and repulsive parts, and a Gent-Schultz dependence of the work of adhesion on crack speed. Their model was used to fit the experimental data, with ad-hoc Gent-Schultz parameters, determined in a way to reproduce the results for different probe diameters.

In contrast, here we started with an independent characterization of the rate-dependent adhesion law under quasi-static conditions (see Fig. \ref{fig:fig5}), i.e., without vibrations. By providing all the dimensional characteristic parameters of our experimental setup as input to the model, it successfully predicted the experimental results, including the time-dependent evolution of the contact radius, spring reaction force and the pull-off force, without any adjustable parameter. Nevertheless, in experiments, the pull-off force saturated above an excitation amplitude of about $100$ $\mu$m. This saturation, which could not be predicted by the model, may be related to the onset of local surface instabilities detected in the contact area snapshot at the pull-off point (see Fig. \ref{fig:fig7}) perhaps due to local circumferential compressive stresses arising within the contact area. Notice that, once the indenter detached from the substrate, we could not identify any permanent damage on the PDMS surface. In order to explain the saturation of the pull-off force, \cite{Shui2020} have proposed a truncation of the Gent and Schultz rate-dependent adhesion law at a critical crack speed $v_C$. This scenario does not find an experimental evidence in our measurements, where the effective work of adhesion was found to monotonically increase with the crack speed (see Fig. \ref{fig:fig5}). {Moreover, the pull-off saturation we found appeared to be frequency dependent (see Fig. \ref{fig:fig6}b), which does not support the existence of a speed cut-off in the Gent and Schultz law, which would have appeared as an inherent property of the interface providing the same pull-off saturation independently on the excitation frequency."}      

In the spirit to clarify the mechanisms of vibroadhesion, providing reference experimental data to the community, we have limited the experiments in the frequency interval $[200, 400]$ Hz where we found results to be robust and repeatable. By considering also equipment limitations, it turned out this interval coincided with a substrate excitation frequency of about $\simeq2.5$ times larger than the expected system resonance $f_{nc}\approx112$ Hz, which the numerical model predicted to be the one with the highest adhesion enhancement. However, testing the system close to resonance appeared difficult to realize in practice due to the noticeable large oscillations of the indenter, which made the tests much more sensitive to small variations of the system parameters. 

{Notice that, we have found that the system response is apparatus-dependent in the range of excitation frequency close to the system resonance, while if $f>>f_{nc}$ there is no dependence on the indenter mass and apparatus stiffness as we discuss in Appendix C.}

\section{Conclusions}\label{sec:conc}
In this work, we considered the problem of a rigid spherical indenter suspended on a compliant spring that makes contact with a soft, viscoelastic adhesive substrate vibrating at a high frequency, $f$, and with a micrometrical amplitude, $X_b$. First, a lumped dynamical model of the nonlinear oscillator in contact with the vibrating substrate was derived. For the contact model, short-range adhesion was assumed, resulting in the classical JKR contact model for an adhesive sphere with an effective, rate-dependent work of adhesion. The model showed that by tuning the vibration amplitude and frequency, the detachment force of the oscillator can be enhanced by more than one order of magnitude compared to the quasi-static pull-off force in the JKR model.

The indenter behaves as a nonlinear oscillator. At a given indentation: (i) for $f$ below the resonance frequency $f_nc$, the substrate moves and the oscillator follows, resulting in minimal variation in indentation and a very weak effect on adhesion; (ii) for $f \simeq f_nc$, close to the system resonance frequency, the oscillator starts vibrating with high amplitude, causing large variations in both indentation and contact radius, which fosters a significant amplification of the pull-off force; (iii) for $f \gg f_nc$, the oscillator oscillates mildly, with the result that the oscillation in indentation is mostly determined by the vibration of the substrate.

Following the insights gathered from the model, an experimental campaign was conducted, in which a borosilicate sphere suspended on a spring was brought into contact with a soft PDMS layer, $3$ mm thick, and excited by high-frequency ($f = [200, 300, 400]$ Hz) micrometrical vibrations ($X_b \in [0, 160]$ $\mu m$). The experimental results qualitatively agreed with the model prediction: as soon as the vibration started, the contact radius increased abruptly, while the average contact force remained almost constant. When unloading started, the curve average contact radius $\overline{a}$ versus average contact force $\overline{F}_c$ followed the JKR contact model, but with an enhanced work of adhesion, resulting in a pull-off force that experimentally we found up to 14.16 times larger than the quasi static case without vibrations.  

To quantitatively compare with the model predictions, we conducted an independent assessment of the rate-dependent adhesion performance of the glass sphere-PDMS substrate interface. The rigid sphere was unloaded from the PDMS substrate at several unloading rates, and the relation between the effective surface energy, $\Delta \gamma_{eff}$, and the crack speed was obtained. During the characterization, the vibration excitation of the substrate was switched off. Having obtained the parameters of the Gent and Schultz adhesion law ($v_0=44.5$ $\mu$m/s and $\alpha=0.531$) for the interface, we exploited the mechanical model to predict the experimental results. This resulted in a very satisfactory comparison, not only for single-value predictions such as the pull-off force, but also for the entire evolution of average contact radius and contact force over time.

We found that model predictions become unreliable above a certain vibration amplitude threshold ($X_b \approx 100$ $\mu$m in our tests), where the experimental pull-off force saturates.  Careful imaging of the contact area evolution at the point of pull-off revealed that, at high vibration amplitudes, surface wrinkles appear in an annular region around the contact area, which may be related to elastic instabilities due to the compressive circumferential stress arising within the contact area, which effect is not included in the theoretical model. How and whether this contributes to the saturation of the pull-off force at high vibration amplitudes remains a subject for further investigation.

In conclusion, we have provided a comprehensive mechanical model to accurately predict adhesion enhancement due to mechanical vibrations without any adjustable parameters. Although we have demonstrated that the model quantitatively agrees with the experimental results, several points still need to be clarified. These include the saturation of the pull-off force, the effect of boundary conditions, the loading-unloading protocol, and the influence of material properties, which will be a matter of further investigation.

\section*{CRediT authorship contribution statement}

Michele Tricarico: Investigation, Methodology, Data curation, Writing – original draft, Writing – review \& editing. Michele Ciavarella: Writing – review \& editing, Validation, Supervision. Antonio Papangelo: Conceptualization, Methodology, Resources, Investigation, Funding acquisition, Supervision, Writing – original draft, Writing – review \& editing,  

\section*{Acknowledgment}

All authors were partly supported by the Italian Ministry of University and Research under the Programme “Department of Excellence” Legge 232/2016 (Grant No. CUP - D93C23000100001). A.P. and M.T. were supported by the European Union (ERC-2021-STG, “Towards Future Interfaces With Tuneable Adhesion By Dynamic Excitation” - SURFACE, Project ID: 101039198, CUP: D95F22000430006). Views and opinions expressed are however those of the authors only and do not necessarily reflect those of the European Union or the European Research Council. Neither the European Union nor the granting authority can be held responsible for them. M.C. was partly supported by the European Union through the program – Next Generation EU (PRIN-2022-PNRR, "Fighting blindness with two photon polymerization of wet adhesive, biomimetic scaffolds for neurosensory REtina-retinal Pigment epitheliAl Interface Regeneration" - REPAIR, Project ID: P2022TTZZF, CUP: D53D23018570001). A.P. is thankful to Julien Scheibert (LTDS, CNRS, Lyon) for insightful discussions and for the support received to set-up the experimental test-rig.   

\section*{Data availability}
The dataset generated for this article is available on Zenodo. 

\appendix
\section{Characterisation of the PDMS sample}
\label{sec:appendixA}

\begin{figure}[H]
\begin{center}
\includegraphics[width=5.5in]{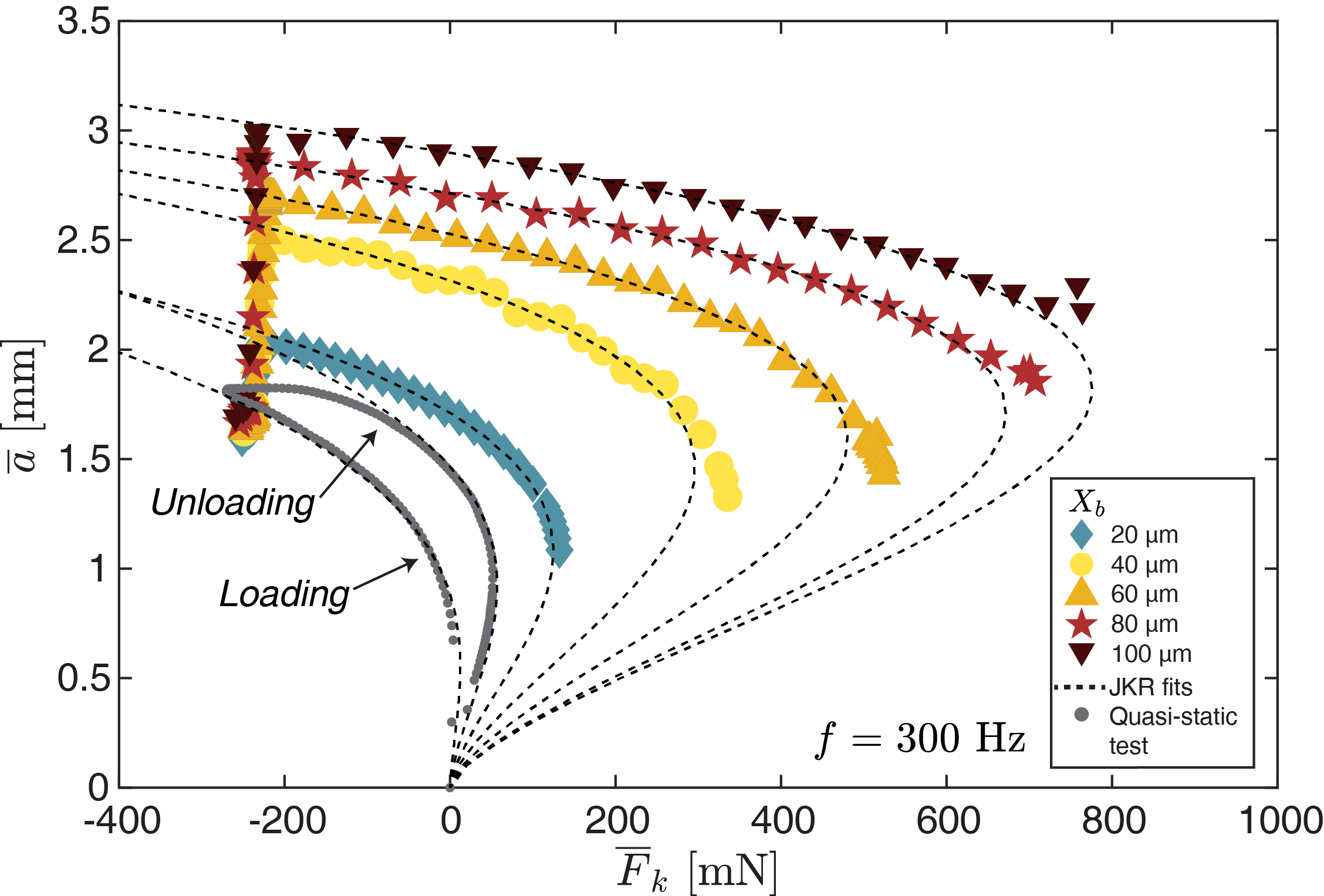}%
\caption{Average spring reaction force ($\overline{F}_{k}$) vs. average contact radius ($\overline{a}$) curves for the quasi-static test (gray dots) and dynamic experimental tests (markers) performed at $f=300$ Hz and $X_b=[20,40,60,80,100]$ $\mu$m with the corresponding JKR fits (black dashed lines). {Notice that the black dashed lines reported should be interpreted as JKR curves with effective properties, while the solid lines in Fig. \ref{fig:fig6}a were obtained upon integration of the full dynamical model (Eq. (\ref{SystEq})).}}
\label{fig:figAppA1}%
\end{center}
\end{figure}

\begin{figure}[H]
\begin{center}
\includegraphics[width=5.5in]{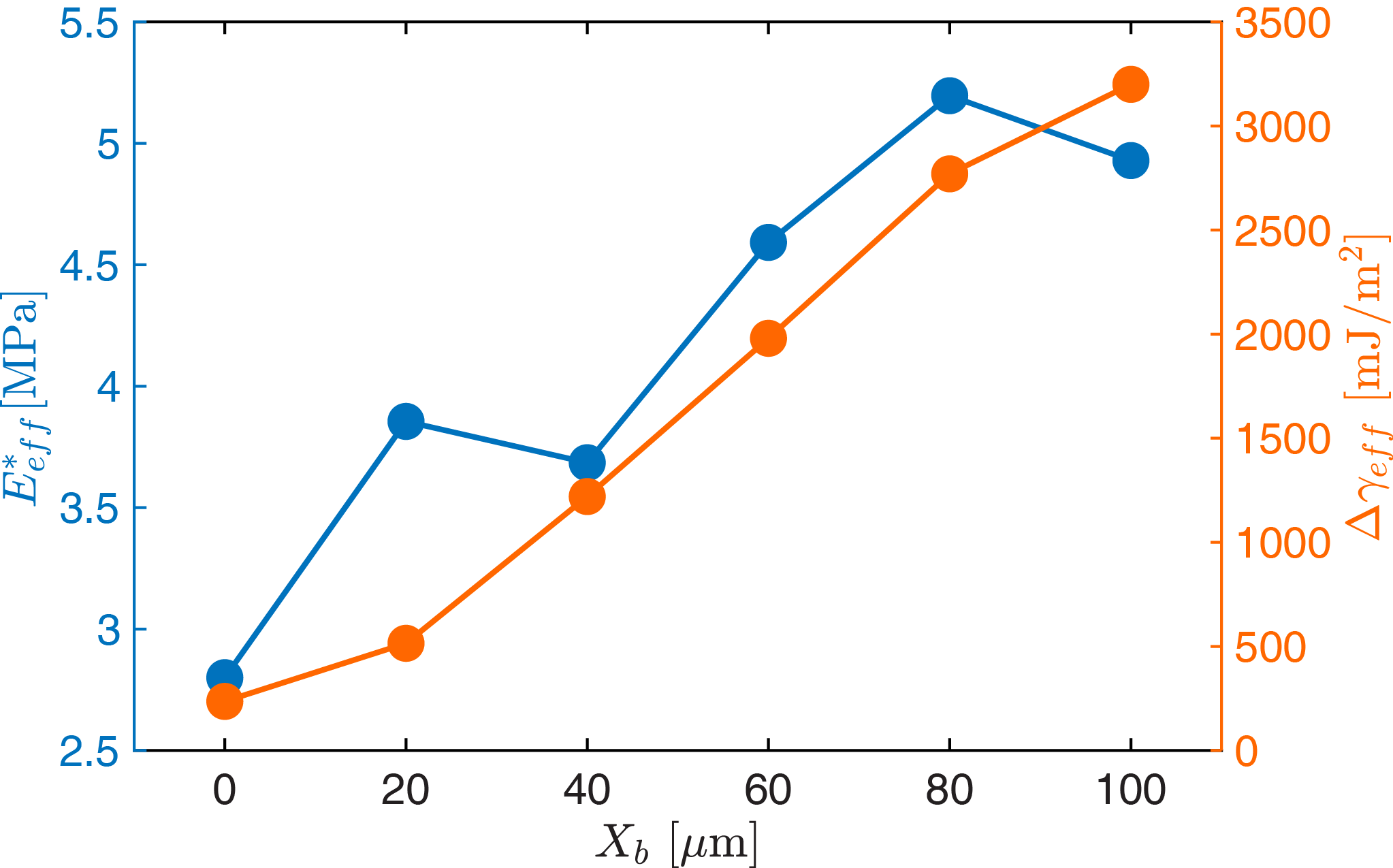}%
\caption{Evolution of the effective contact modulus $E^*_{eff}$ 
 and effective work of adhesion $\Delta \gamma_{eff}$ as functions of the vibration amplitude $X_b$ ($f=300$ Hz). The data points were obtained by fitting the experimental force vs. contact radius curves at unloading with a JKR model, as shown in Fig. \ref{fig:figAppA1}} 
\label{fig:figAppA2}%
\end{center}
\end{figure}

Static (i.e. without vibration) and dynamic (vibroadhesive) tests were performed to characterize the mechanical response of the PDMS sample.
We used the same preload $F_{c0}$=-250 mN for all the other tests. 
By fitting the quasi-static test at unloading (gray markers in Fig. \ref{fig:figAppA1}) with JKR curves, we found $\Delta \gamma_0 = 236$  $\text{mJ/m}^2$ and $E_{0}^{\ast}=2.8$ MPa. 
Similarly, also the dynamic tests were fitted with "effective" JKR curves. We found that both $\Delta \gamma_{eff}$ and $E_{eff}^{\ast}$ increased with $X_b$ (Figure \ref{fig:figAppA2}), due to the viscoelastic nature of the substrate.
In particular, $E_{eff}^{\ast}$ represents an "effective contact modulus" for the effective JKR model. {The larger the vibration amplitude, the larger the volume of the material excited, which yields an effective stiffening of the bulk. Indeed, below the indenter there exists a volume of material $V\approx \pi a^2 \delta$ that stiffens due to the indentation cycle, which is surrounded by a relaxed material, which modulus is $\sim E_0$. The effective elastic modulus must take all this into account and it results in an effective stiffening when $X_b$, hence $V$, increases. Notice that in our experiments the maximum ratio $a/t$ never exceeded 1, hence finite size effects should be considered minimal on the reported results (see \cite{bentall1968elastic,perriot2004elastic,maghami2024viscoelastic}).}
As input for the numerical model, we used an average value of $E_{eff}^{\ast}=4.5$ MPa (see Fig. \ref{fig:figAppA2}). {By using Eq. (9) in \cite{Maghami2024} for the complex compliance $\overline{C}\left(  \omega\right)$ and considering the complex moduls  $\overline{E}\left(  \omega\right)=1/\overline{C}\left(  \omega\right)$, one finds at $f=300$ Hz a factor $|\overline{E}\left(  2\pi f \right)|/E_0 \simeq 1.9$, which is very close to the stiffening factor used here, i.e. $E_{eff}/E_0\simeq1.6$. Clearly, one should not expect the DMA stiffening factor to plainly apply to our contact problem. The DMA test is performed as a uniaxial tensile test, with an oscillatory strain of $\epsilon=0.1\%$, which clearly does not correspond to the multiaxial contact problem considered here. Modeling this effect is not straightforward, since it would require a dedicate finite element numerical analysis that takes into account the layer thickness, the viscoelastic nature of PDMS and its hyperelastic behavior, which goes beyond the scope of the present work.}

\section{Comparison with Shui et al. (2020)}
\label{sec:appendixB}

\begin{figure}[t]
\begin{center}
\includegraphics[width=4.5in]{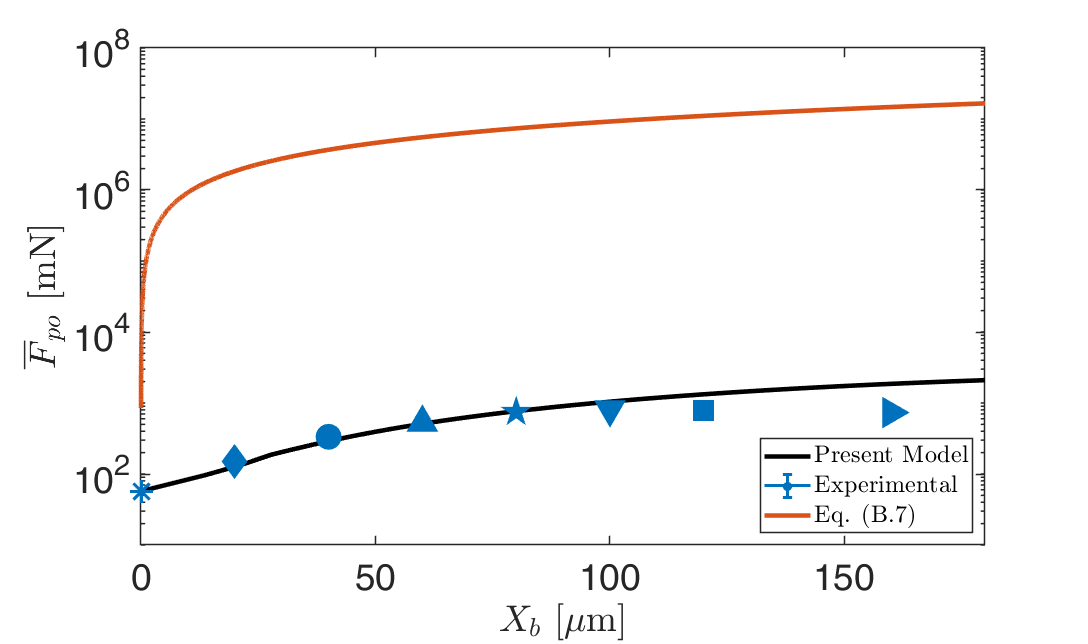}%
\caption{{Pull-off force as a function of the vibration amplitude $X_b$. The markers represent the experimental data, the present model is shown as a black solid line while the predictions of Eq. (\ref{ShuiFpo}) by \cite{Shui2020} is shown as an orange solid line. The case shown corresponds to $f= 300$ Hz.}}
\label{fig:figAppB10}%
\end{center}
\end{figure}

Shui et al. (2020) has proposed a slightly different model for the system in
Fig. \ref{fig:fig1} starting from the hypothesis that the effect of the vibrating
substrate can be taken into account by a forcing term $f(t)\simeq-m\omega
^{2}X_{b}\sin\left(  \omega t\right)  $. {If, in the model proposed here, we write the indentation as a function of the oscillator and base position}

\begin{align}
\delta &  =x+x_{b}=x+X_{b}\sin\left(  \omega t\right) \\
\overset{\cdot}{\delta}  &  =\overset{\cdot}{x}+\overset{\cdot}{x}%
_{b}=\overset{\cdot}{x}+\omega X_{b}\cos\left(  \omega t\right) \\
\overset{\cdot\cdot}{\delta}  &  =\overset{\cdot\cdot}{x}+\overset{\cdot\cdot
}{x}_{b}=\overset{\cdot\cdot}{x}-\omega^{2}X_{b}\sin\left(  \omega t\right)
\end{align}
and substituting into the equilibrium equation Eq. (\ref{Equi}), one obtains%

\begin{equation}
m\overset{\cdot\cdot}{\delta}+c\overset{\cdot}{\delta}+k(\delta-x_{k}%
)=-m\omega^{2}X_{b}\sin\left(  \omega t\right)  +c\omega X_{b}\cos\left(
\omega t\right)  +kX_{b}\sin\left(  \omega t\right)  +F_{c}-F_{c,\delta=0}%
\end{equation}
{From the JKR model writing the contact force} $F_{c}=\frac{Ka}{2}\left(  \frac{a^{2}}{R}%
-3\delta\right)  $, with $K=4/3E^*$, and considering $k\rightarrow0$ and $c\ll m\omega$, one obtains%

\begin{equation}
\frac{Ka}{2}\left(  3\delta-\frac{a^{2}}{R}\right)  +m\overset{\cdot\cdot
}{\delta}+c\overset{\cdot}{\delta}=-m\omega^{2}X_{b}\sin\left(  \omega
t\right)  \label{EqShui}%
\end{equation}
which, apart from the convention of the sign, coincides with the equilibrium
equation in Shui et al. (2020). Hence, the model presented here, generalizes
that presented by Shui et al. (2020), which broadly speaking would correspond
to the "force controlled" scenario. Starting from (\ref{EqShui}), Shui et al.
(2020) have developed an approximate solution for the average contact force
and average contact radius which in our dimensionless notation reads%

\begin{equation}
\overline{\widetilde{F}}=-\overline{\widetilde{a}}^{3}+\sqrt{6\overline
{\widetilde{a}}^{3}}+\frac{3}{2}\frac{\widetilde{f}^{2}\widetilde{X}%
_{b}\overline{\widetilde{a}}}{\sqrt{\left(  \frac{3}{2\widetilde{k}}%
\overline{\widetilde{a}}-\widetilde{f}^{2}\right)  ^{2}+\left(  2\zeta
\widetilde{f}\right)  ^{2}}};\qquad\omega\gg0,v<v_{c}\label{ShuiF}%
\end{equation}
where the symbols $\left\{  \overline{\widetilde{F}},\overline{\widetilde{a}%
}\right\}  $ stand respectively for the average contact force and average
contact radius, hence the pull-off force is the maximum of $\overline{F}$,
which, in the supplementary information, Shui et al (2020)\ estimates as%

\begin{equation}
\overline{\widetilde{F}}_{po}=\left\{
\begin{array}
[c]{cc}%
3/2 & \qquad\widetilde{X}_{b}\rightarrow0\\
\widetilde{X}_{b}\widetilde{k}\widetilde{f}^{2}\sqrt{1+\left(  \frac
{\widetilde{f}}{2\zeta}\right)  ^{2}} & \qquad\widetilde{X}_{b}>>0
\end{array}
\right.  \label{ShuiFpo}%
\end{equation}
Nevertheless, the predictions of Eq. (\ref{ShuiFpo}) compared very poorly with
our experimental data (see Fig. \ref{fig:figAppB10}). {The striking difference between the model proposed here and the approximate solution derived by \cite{Shui2020} is in the definition of the damping coefficient $c$. In the model proposed here $c$ includes the dissipative phenomena at play when the indenter is freely oscillating (not in contact), while the dissipative contribution coming from the viscoelastic bulk have been independently characterized by the Gent and Schultz law. Conversely, the model proposed by \cite{Shui2020}, incorporates within the damping coefficient $c$ all the dissipative phenomena of the system, including the hysteresis originated from the viscoelastic nature of the bulk material. In the supplementary information of their paper, \cite{Shui2020} suggest that the damping coefficient should be determined from the input power provided to the system, nevertheless it remains unclear how this should be related to the rate-dependent adhesive behavior of the soft interface.}

Going back to the system equilibrium of motion in the limit of force control
(Eq. (\ref{EqShui})), if one considers small vibrations around an equilibrium
position $\delta=\overline{\delta}+\delta_{A}\sin\left(  \omega t\right)  $
and $a=\overline{a}+a_{A}\sin\left(  \omega t\right)  $, assuming $\delta
_{A}<<\overline{\delta}$ and $a_{A}<<\overline{a}$ and linearizing around the
equilibrium position, the indentation amplitude $\delta_{A}$ would read%

\begin{equation}
\left\vert \frac{\delta_{A}}{X_{b}}\right\vert =\left\vert \frac{\left(
\omega/\omega_{nc}\right)  ^{2}}{\sqrt{\left(  1-\left(  \omega/\omega
_{nc}\right)  ^{2}\right)  ^{2}+4\zeta^{2}\left(  \omega/\omega_{nc}\right)
^{2}}}\right\vert \label{Eq:app:B8}
\end{equation}
being $\omega_{nc}=\sqrt{\left(  \left.  \frac{dF_{c}}{d\delta}\right\vert
_{\delta=\overline{\delta}}\right)  /m}$ which gives the "resonance" frequency
of the system. Notice that the latter is $\overline{\delta}$ dependent. {Equation (\ref{Eq:app:B8}) is only a simple linearized approximation of the system dynamical behavior. A full treatment of the nonlinear dynamical system response is out of the scope of the present work.}

\section{High frequency model}
\label{sec:appendixC}

In Fig. \ref{fig:fig2} it was shown that, for very large excitation frequency $\widetilde{f}\gg\widetilde{f}_{nc}$, the oscillations in the spherical probe indentation will be fully determined by the substrate vibration. In this case it is possible to further simplify the governing equations of the system by considering that $\widetilde{x}\approx0$, which implies the oscillator dynamics can be neglected and the system governing equations simplify as 
\begin{equation}
\begin{array}
[c]{l}%
\widetilde{a}^{\prime}=\left\{
\begin{array}
[c]{lc}%
-\widetilde{v}_{0}\left(  \frac{3}{8\widetilde{a}}\left(  \widetilde{a}%
^{2}-\widetilde{\delta}  \right)  ^{2}-1\right)  ^{1/\alpha} & \qquad\text{, if }%
\widetilde{a}^{\prime}<0\\
\widetilde{v}_{0}\left(  \frac{8\widetilde{a}}{3}\left(  \widetilde{a}%
^{2}-\widetilde{\delta}  \right)^{-2}-1 \right)^{1/\alpha} & \qquad\text{, if
}\widetilde{a}^{\prime}>0
\end{array}
\right.
\end{array}\label{SystEqHighFreq}%
\end{equation}

where $\widetilde{\delta} = \widetilde{\delta}_0 + \widetilde{r}\widetilde{t} + \widetilde{X}_b \sin(\widetilde{f}\widetilde{t})$, with $\widetilde{\delta}_0$ representing the initial indentation at a certain preload giving  $\widetilde{a}_0$ as the corresponding initial contact radius according to the "quasi-static" JKR model. The ODE function in Eq. (\ref{SystEqHighFreq}) can be easily integrated, yielding the contact radius evolution $\widetilde{a}(\widetilde{t})$ and, substituting into Eq. (\ref{JKRdlessFc}), the contact force. We refer to this as the "high-frequency" model. 

\begin{figure}[h!]
\begin{center}
\includegraphics[width=4.5in]{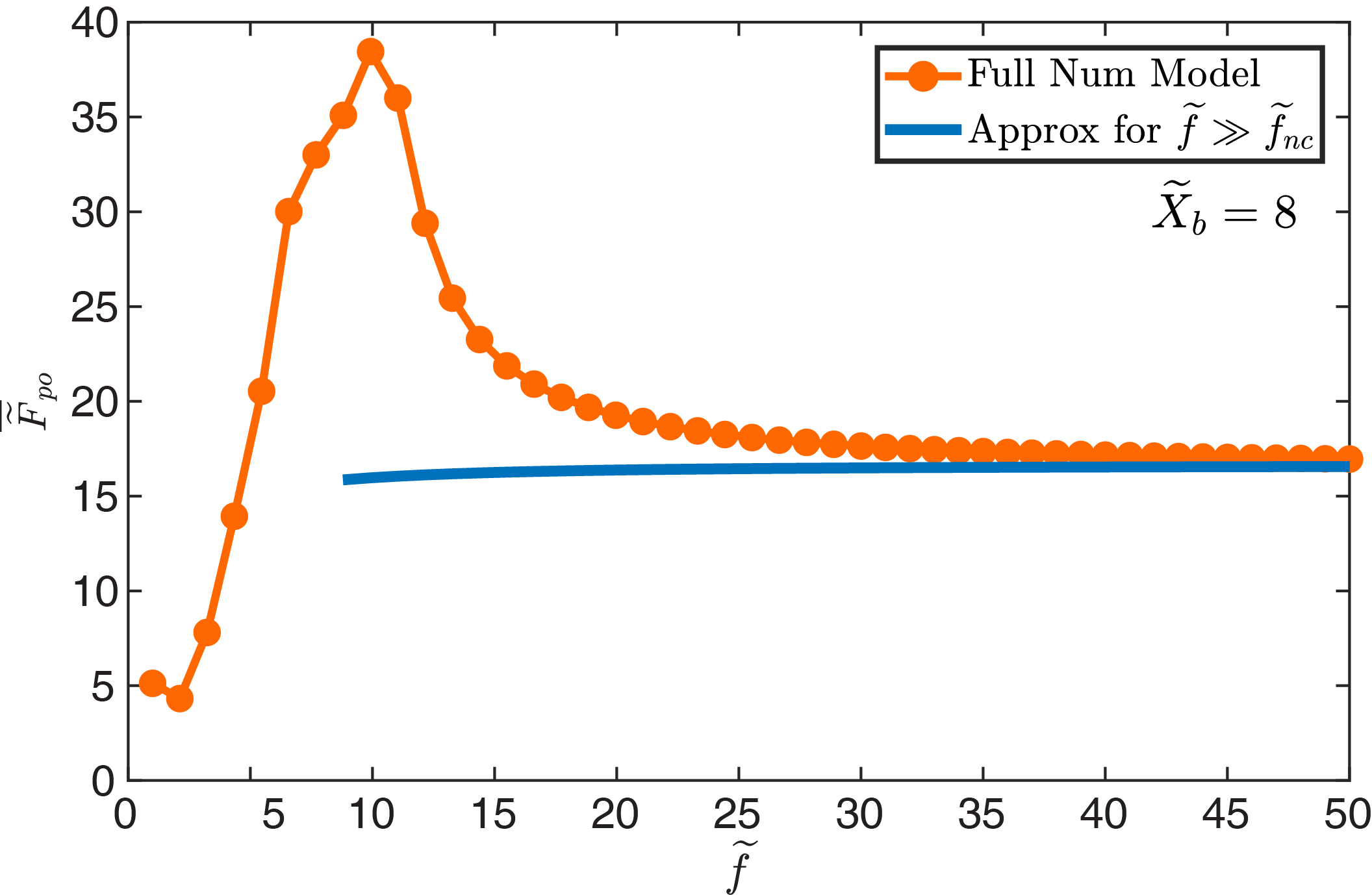}%
\caption{Comparison between the full numerical model (orange line) and the "high-frequency" model (blue line, Eq. (\ref{SystEqHighFreq})). The predictions of the dimensionless pull-off force $\overline{\widetilde{F}}_{po}$ as a function of the dimensionless excitation frequency $\widetilde{f}$ for the two models converge for large values of frequency. {Here the dimensionless amplitude was fixed to $\widetilde{X}_b=8$ and the unloading rate to $\widetilde{r}=-0.1$, which gives already a certain pull-off enhancement with respect to the quasi-static JKR case $\widetilde{F}_{JKR}=3/2$ at $\widetilde{f}=0$.}}
\label{fig:figApp}%
\end{center}
\end{figure}

Note that ignoring the oscillator dynamics is a valid approximation only in the high-frequency regime. Figure \ref{fig:figApp} compares the dimensionless pull-off force $\overline{\widetilde{F}}_{po}$ as a function of the dimensionless excitation frequency $\widetilde{f}$, obtained by integrating the full numerical model (Eq. (\ref{SystEq}), red curve) and the approximate high-frequency model (Eq. \ref{SystEqHighFreq}, blue curve). The high-frequency model significantly underestimates the interfacial adhesion strength near resonance ($\widetilde{f} \approx 10$), as it neglects the contribution of the oscillator dynamics in the sphere indentation. However, as the frequency increases, this contribution diminishes, resulting in an accurate estimation of the pull-off force without requiring integration of the full system.

\clearpage
\bibliographystyle{elsarticle-harv} 
\bibliography{cas-refs}

\begin{thebibliography}{66}
\expandafter\ifx\csname natexlab\endcsname\relax\def\natexlab#1{#1}\fi
\providecommand{\url}[1]{\texttt{#1}}
\providecommand{\href}[2]{#2}
\providecommand{\path}[1]{#1}
\providecommand{\DOIprefix}{doi:}
\providecommand{\ArXivprefix}{arXiv:}
\providecommand{\URLprefix}{URL: }
\providecommand{\Pubmedprefix}{pmid:}
\providecommand{\doi}[1]{\href{http://dx.doi.org/#1}{\path{#1}}}
\providecommand{\Pubmed}[1]{\href{pmid:#1}{\path{#1}}}
\providecommand{\bibinfo}[2]{#2}
\ifx\xfnm\relax \def\xfnm[#1]{\unskip,\space#1}\fi
\bibitem[{Afferrante and Violano(2022)}]{afferrante2022effective}
\bibinfo{author}{Afferrante, L.}, \bibinfo{author}{Violano, G.}, \bibinfo{year}{2022}.
\newblock \bibinfo{title}{On the effective surface energy in viscoelastic hertzian contacts}.
\newblock \bibinfo{journal}{Journal of the Mechanics and Physics of Solids} \bibinfo{volume}{158}, \bibinfo{pages}{104669}.
\bibitem[{Argatov et~al.(2024)Argatov, Papangelo and Ciavarella}]{Argatov2024}
\bibinfo{author}{Argatov, I.}, \bibinfo{author}{Papangelo, A.}, \bibinfo{author}{Ciavarella, M.}, \bibinfo{year}{2024}.
\newblock \bibinfo{title}{Vibroadhesion - an asymptotic model}.
\newblock \bibinfo{journal}{submitted} .
\bibitem[{Arzt et~al.(2021)Arzt, Quan, McMeeking and Hensel}]{arzt2021functional}
\bibinfo{author}{Arzt, E.}, \bibinfo{author}{Quan, H.}, \bibinfo{author}{McMeeking, R.M.}, \bibinfo{author}{Hensel, R.}, \bibinfo{year}{2021}.
\newblock \bibinfo{title}{Functional surface microstructures inspired by nature--from adhesion and wetting principles to sustainable new devices}.
\newblock \bibinfo{journal}{Progress in Materials Science} \bibinfo{volume}{120}, \bibinfo{pages}{100823}.
\bibitem[{Autumn and Peattie(2002)}]{autumn2002mechanisms}
\bibinfo{author}{Autumn, K.}, \bibinfo{author}{Peattie, A.M.}, \bibinfo{year}{2002}.
\newblock \bibinfo{title}{Mechanisms of adhesion in geckos}.
\newblock \bibinfo{journal}{Integrative and comparative biology} \bibinfo{volume}{42}, \bibinfo{pages}{1081--1090}.
\bibitem[{Barthel(2024)}]{barthel2024linear}
\bibinfo{author}{Barthel, E.}, \bibinfo{year}{2024}.
\newblock \bibinfo{title}{The linear viscoelastic fracture theory applies to soft solids better when they are… viscoelastic}.
\newblock \bibinfo{journal}{Proceedings of the Royal Society A} \bibinfo{volume}{480}, \bibinfo{pages}{20230561}.
\bibitem[{Bentall and Johnson(1968)}]{bentall1968elastic}
\bibinfo{author}{Bentall, R.}, \bibinfo{author}{Johnson, K.}, \bibinfo{year}{1968}.
\newblock \bibinfo{title}{An elastic strip in plane rolling contact}.
\newblock \bibinfo{journal}{International Journal of Mechanical Sciences} \bibinfo{volume}{10}, \bibinfo{pages}{637--663}.
\bibitem[{Cacucciolo et~al.(2022)Cacucciolo, Shea and Carbone}]{Cacucciolo2022}
\bibinfo{author}{Cacucciolo, V.}, \bibinfo{author}{Shea, H.}, \bibinfo{author}{Carbone, G.}, \bibinfo{year}{2022}.
\newblock \bibinfo{title}{Peeling in electroadhesion soft grippers}.
\newblock \bibinfo{journal}{Extreme Mechanics Letters} \bibinfo{volume}{50}, \bibinfo{pages}{101529}.
\bibitem[{Carbone et~al.(2022)Carbone, Mandriota and Menga}]{carbone2022theory}
\bibinfo{author}{Carbone, G.}, \bibinfo{author}{Mandriota, C.}, \bibinfo{author}{Menga, N.}, \bibinfo{year}{2022}.
\newblock \bibinfo{title}{Theory of viscoelastic adhesion and friction}.
\newblock \bibinfo{journal}{Extreme Mechanics Letters} \bibinfo{volume}{56}, \bibinfo{pages}{101877}.
\bibitem[{Chen et~al.(2020)Chen, Cao, Sarparast, Yuan, Dong, Tan and Cao}]{chen2020soft}
\bibinfo{author}{Chen, S.}, \bibinfo{author}{Cao, Y.}, \bibinfo{author}{Sarparast, M.}, \bibinfo{author}{Yuan, H.}, \bibinfo{author}{Dong, L.}, \bibinfo{author}{Tan, X.}, \bibinfo{author}{Cao, C.}, \bibinfo{year}{2020}.
\newblock \bibinfo{title}{Soft crawling robots: design, actuation, and locomotion}.
\newblock \bibinfo{journal}{Advanced Materials Technologies} \bibinfo{volume}{5}, \bibinfo{pages}{1900837}.
\bibitem[{Ciavarella et~al.(2021)Ciavarella, Cricr{\`\i} and McMeeking}]{ciavarella2021comparison}
\bibinfo{author}{Ciavarella, M.}, \bibinfo{author}{Cricr{\`\i}, G.}, \bibinfo{author}{McMeeking, R.}, \bibinfo{year}{2021}.
\newblock \bibinfo{title}{A comparison of crack propagation theories in viscoelastic materials}.
\newblock \bibinfo{journal}{Theoretical and applied fracture mechanics} \bibinfo{volume}{116}, \bibinfo{pages}{103113}.
\bibitem[{Ciavarella et~al.(2024)Ciavarella, Tricarico and Papangelo}]{CiaTricPap2024}
\bibinfo{author}{Ciavarella, M.}, \bibinfo{author}{Tricarico, M.}, \bibinfo{author}{Papangelo, A.}, \bibinfo{year}{2024}.
\newblock \bibinfo{title}{On the dynamic jkr adhesion problem}.
\newblock \bibinfo{journal}{submitted} \DOIprefix\doi{10.13140/RG.2.2.14211.31526}.
\bibitem[{Creton and Ciccotti(2016)}]{Creton2016}
\bibinfo{author}{Creton, C.}, \bibinfo{author}{Ciccotti, M.}, \bibinfo{year}{2016}.
\newblock \bibinfo{title}{Fracture and adhesion of soft materials: a review}.
\newblock \bibinfo{journal}{Reports on Progress in Physics} \bibinfo{volume}{79}, \bibinfo{pages}{046601}.
\newblock \URLprefix \url{https://dx.doi.org/10.1088/0034-4885/79/4/046601}, \DOIprefix\doi{10.1088/0034-4885/79/4/046601}.
\bibitem[{Delplanque et~al.(2022)Delplanque, Aymard, Dalmas and Scheibert}]{delplanque2022solving}
\bibinfo{author}{Delplanque, E.}, \bibinfo{author}{Aymard, A.}, \bibinfo{author}{Dalmas, D.}, \bibinfo{author}{Scheibert, J.}, \bibinfo{year}{2022}.
\newblock \bibinfo{title}{Solving curing-protocol-dependent shape errors in pdms replication}.
\newblock \bibinfo{journal}{Journal of Micromechanics and Microengineering} \bibinfo{volume}{32}, \bibinfo{pages}{045006}.
\bibitem[{Dorogin and Persson(2018)}]{dorogin2018contact}
\bibinfo{author}{Dorogin, L.}, \bibinfo{author}{Persson, B.}, \bibinfo{year}{2018}.
\newblock \bibinfo{title}{Contact mechanics for polydimethylsiloxane: from liquid to solid}.
\newblock \bibinfo{journal}{Soft matter} \bibinfo{volume}{14}, \bibinfo{pages}{1142--1148}.
\bibitem[{Duan et~al.(2023)Duan, Yu, Cui, Zhang, Zhang and Tian}]{Duan2023}
\bibinfo{author}{Duan, W.}, \bibinfo{author}{Yu, Z.}, \bibinfo{author}{Cui, W.}, \bibinfo{author}{Zhang, Z.}, \bibinfo{author}{Zhang, W.}, \bibinfo{author}{Tian, Y.}, \bibinfo{year}{2023}.
\newblock \bibinfo{title}{Bio-inspired switchable soft adhesion for the boost of adhesive surfaces and robotics applications: A brief review}.
\newblock \bibinfo{journal}{Advances in colloid and interface science} \bibinfo{volume}{313}, \bibinfo{pages}{102862}.
\bibitem[{Edsinger and Kemp(2007)}]{edsinger2007human}
\bibinfo{author}{Edsinger, A.}, \bibinfo{author}{Kemp, C.C.}, \bibinfo{year}{2007}.
\newblock \bibinfo{title}{Human-robot interaction for cooperative manipulation: Handing objects to one another}, in: \bibinfo{booktitle}{RO-MAN 2007-The 16th IEEE International Symposium on Robot and Human Interactive Communication}, \bibinfo{organization}{IEEE}. pp. \bibinfo{pages}{1167--1172}.
\bibitem[{Gent and Schultz(1972)}]{gent1972effect}
\bibinfo{author}{Gent, A.}, \bibinfo{author}{Schultz, J.}, \bibinfo{year}{1972}.
\newblock \bibinfo{title}{Effect of wetting liquids on the strength of adhesion of viscoelastic material}.
\newblock \bibinfo{journal}{The Journal of Adhesion} \bibinfo{volume}{3}, \bibinfo{pages}{281--294}.
\bibitem[{Giordano et~al.(2024)Giordano, Scharff, Carlotti, Gagliardi, Filippeschi, Mondini, Papangelo and Mazzolai}]{giordano2024mechanochromic}
\bibinfo{author}{Giordano, G.}, \bibinfo{author}{Scharff, R.B.N.}, \bibinfo{author}{Carlotti, M.}, \bibinfo{author}{Gagliardi, M.}, \bibinfo{author}{Filippeschi, C.}, \bibinfo{author}{Mondini, A.}, \bibinfo{author}{Papangelo, A.}, \bibinfo{author}{Mazzolai, B.}, \bibinfo{year}{2024}.
\newblock \bibinfo{title}{Mechanochromic suction cups for local stress detection in soft robotics}.
\newblock \bibinfo{journal}{Advanced Intelligent Systems} , \bibinfo{pages}{2400254}.
\bibitem[{Gorb et~al.(2007)Gorb, Varenberg, Peressadko and Tuma}]{gorb2007biomimetic}
\bibinfo{author}{Gorb, S.}, \bibinfo{author}{Varenberg, M.}, \bibinfo{author}{Peressadko, A.}, \bibinfo{author}{Tuma, J.}, \bibinfo{year}{2007}.
\newblock \bibinfo{title}{Biomimetic mushroom-shaped fibrillar adhesive microstructure}.
\newblock \bibinfo{journal}{Journal of the Royal Society Interface} \bibinfo{volume}{4}, \bibinfo{pages}{271--275}.
\bibitem[{Greenwood(2004)}]{greenwood2004theory}
\bibinfo{author}{Greenwood, J.}, \bibinfo{year}{2004}.
\newblock \bibinfo{title}{The theory of viscoelastic crack propagation and healing}.
\newblock \bibinfo{journal}{Journal of Physics D: Applied Physics} \bibinfo{volume}{37}, \bibinfo{pages}{2557}.
\bibitem[{Greenwood and Johnson(1981)}]{greenwood1981mechanics}
\bibinfo{author}{Greenwood, J.}, \bibinfo{author}{Johnson, K.}, \bibinfo{year}{1981}.
\newblock \bibinfo{title}{The mechanics of adhesion of viscoelastic solids}.
\newblock \bibinfo{journal}{Philosophical Magazine A} \bibinfo{volume}{43}, \bibinfo{pages}{697--711}.
\bibitem[{Greenwood and Johnson(2006)}]{greenwood2006oscillatory}
\bibinfo{author}{Greenwood, J.}, \bibinfo{author}{Johnson, K.}, \bibinfo{year}{2006}.
\newblock \bibinfo{title}{Oscillatory loading of a viscoelastic adhesive contact}.
\newblock \bibinfo{journal}{Journal of colloid and interface science} \bibinfo{volume}{296}, \bibinfo{pages}{284--291}.
\bibitem[{He et~al.(2022)He, Yang and Jiang}]{he2022modeling}
\bibinfo{author}{He, Z.}, \bibinfo{author}{Yang, Y.}, \bibinfo{author}{Jiang, H.}, \bibinfo{year}{2022}.
\newblock \bibinfo{title}{Modeling interfacial instability patterns during debonding a rigid spherical indenter from thin elastic films}.
\newblock \bibinfo{journal}{Journal of the Mechanics and Physics of Solids} \bibinfo{volume}{169}, \bibinfo{pages}{105089}.
\bibitem[{Huang et~al.(2024)Huang, Liu, Zhu, Jiang, Sun, Yang, Qi and Wang}]{Huang2024}
\bibinfo{author}{Huang, Z.j.}, \bibinfo{author}{Liu, Y.l.}, \bibinfo{author}{Zhu, T.y.}, \bibinfo{author}{Jiang, W.j.}, \bibinfo{author}{Sun, D.x.}, \bibinfo{author}{Yang, J.h.}, \bibinfo{author}{Qi, X.d.}, \bibinfo{author}{Wang, Y.}, \bibinfo{year}{2024}.
\newblock \bibinfo{title}{Phase change hydrogels with tunable adhesion for wearable thermal management and intelligent healthcare}.
\newblock \bibinfo{journal}{Journal of Energy Storage} \bibinfo{volume}{98}, \bibinfo{pages}{113043}.
\bibitem[{Hui et~al.(2022)Hui, Zhu and Long}]{hui2022steady}
\bibinfo{author}{Hui, C.Y.}, \bibinfo{author}{Zhu, B.}, \bibinfo{author}{Long, R.}, \bibinfo{year}{2022}.
\newblock \bibinfo{title}{Steady state crack growth in viscoelastic solids: A comparative study}.
\newblock \bibinfo{journal}{Journal of the Mechanics and Physics of Solids} \bibinfo{volume}{159}, \bibinfo{pages}{104748}.
\bibitem[{Johnson et~al.(1971)Johnson, Kendall and Roberts}]{johnson1971surface}
\bibinfo{author}{Johnson, K.L.}, \bibinfo{author}{Kendall, K.}, \bibinfo{author}{Roberts, A.}, \bibinfo{year}{1971}.
\newblock \bibinfo{title}{Surface energy and the contact of elastic solids}.
\newblock \bibinfo{journal}{Proceedings of the royal society of London. A. mathematical and physical sciences} \bibinfo{volume}{324}, \bibinfo{pages}{301--313}.
\bibitem[{Kamperman et~al.(2010)Kamperman, Kroner, Del~Campo, McMeeking and Arzt}]{kamperman2010functional}
\bibinfo{author}{Kamperman, M.}, \bibinfo{author}{Kroner, E.}, \bibinfo{author}{Del~Campo, A.}, \bibinfo{author}{McMeeking, R.M.}, \bibinfo{author}{Arzt, E.}, \bibinfo{year}{2010}.
\newblock \bibinfo{title}{Functional adhesive surfaces with “gecko” effect: The concept of contact splitting}.
\newblock \bibinfo{journal}{Advanced Engineering Materials} \bibinfo{volume}{12}, \bibinfo{pages}{335--348}.
\bibitem[{Li et~al.(2024)Li, Zhou, Ouyang, Guan, Shen, Saiz, Li and Hou}]{li2024harnessing}
\bibinfo{author}{Li, W.}, \bibinfo{author}{Zhou, R.}, \bibinfo{author}{Ouyang, Y.}, \bibinfo{author}{Guan, Q.}, \bibinfo{author}{Shen, Y.}, \bibinfo{author}{Saiz, E.}, \bibinfo{author}{Li, M.}, \bibinfo{author}{Hou, X.}, \bibinfo{year}{2024}.
\newblock \bibinfo{title}{Harnessing biomimicry for controlled adhesion on material surfaces}.
\newblock \bibinfo{journal}{Small} , \bibinfo{pages}{2401859}.
\bibitem[{Li et~al.(2022)Li, Bai, Li, Li, Li, Lu, Ma, Meng and Tian}]{li2022robust}
\bibinfo{author}{Li, X.}, \bibinfo{author}{Bai, P.}, \bibinfo{author}{Li, X.}, \bibinfo{author}{Li, L.}, \bibinfo{author}{Li, Y.}, \bibinfo{author}{Lu, H.}, \bibinfo{author}{Ma, L.}, \bibinfo{author}{Meng, Y.}, \bibinfo{author}{Tian, Y.}, \bibinfo{year}{2022}.
\newblock \bibinfo{title}{Robust scalable reversible strong adhesion by gecko-inspired composite design}.
\newblock \bibinfo{journal}{Friction} \bibinfo{volume}{10}, \bibinfo{pages}{1192--1207}.
\bibitem[{Linghu et~al.(2024)Linghu, Liu, Yang, Li, Tan, Mohamed, Mohammad, Du, Su, Li et~al.}]{linghu2024fibrillar}
\bibinfo{author}{Linghu, C.}, \bibinfo{author}{Liu, Y.}, \bibinfo{author}{Yang, X.}, \bibinfo{author}{Li, D.}, \bibinfo{author}{Tan, Y.Y.}, \bibinfo{author}{Mohamed, H.B.M.H.}, \bibinfo{author}{Mohammad, F.B.R.}, \bibinfo{author}{Du, Z.}, \bibinfo{author}{Su, J.}, \bibinfo{author}{Li, Y.}, et~al., \bibinfo{year}{2024}.
\newblock \bibinfo{title}{Fibrillar adhesives with unprecedented adhesion strength, switchability and scalability}.
\newblock \bibinfo{journal}{National Science Review} , \bibinfo{pages}{nwae106}.
\bibitem[{Liu et~al.(2023)Liu, Yin, Zhu, Zhao, Yu, Qu and Yang}]{Liu2023}
\bibinfo{author}{Liu, B.}, \bibinfo{author}{Yin, T.}, \bibinfo{author}{Zhu, J.}, \bibinfo{author}{Zhao, D.}, \bibinfo{author}{Yu, H.}, \bibinfo{author}{Qu, S.}, \bibinfo{author}{Yang, W.}, \bibinfo{year}{2023}.
\newblock \bibinfo{title}{Tough and fatigue-resistant polymer networks by crack tip softening}.
\newblock \bibinfo{journal}{Proceedings of the National Academy of Sciences} \bibinfo{volume}{120}, \bibinfo{pages}{e2217781120}.
\bibitem[{Maghami et~al.(2024a)Maghami, Tricarico, Ciavarella and Papangelo}]{maghami2024viscoelastic}
\bibinfo{author}{Maghami, A.}, \bibinfo{author}{Tricarico, M.}, \bibinfo{author}{Ciavarella, M.}, \bibinfo{author}{Papangelo, A.}, \bibinfo{year}{2024}a.
\newblock \bibinfo{title}{Viscoelastic amplification of the pull-off stress in the detachment of a rigid flat punch from an adhesive soft viscoelastic layer}.
\newblock \bibinfo{journal}{Engineering Fracture Mechanics} \bibinfo{volume}{298}, \bibinfo{pages}{109898}.
\bibitem[{Maghami et~al.(2024b)Maghami, Wang, Tricarico, Ciavarella, Li and Papangelo}]{Maghami2024}
\bibinfo{author}{Maghami, A.}, \bibinfo{author}{Wang, Q.}, \bibinfo{author}{Tricarico, M.}, \bibinfo{author}{Ciavarella, M.}, \bibinfo{author}{Li, Q.}, \bibinfo{author}{Papangelo, A.}, \bibinfo{year}{2024}b.
\newblock \bibinfo{title}{Bulk and fracture process zone contribution to the rate-dependent adhesion amplification in viscoelastic broad-band materials}.
\newblock \bibinfo{journal}{Journal of the Mechanics and Physics of Solids} \bibinfo{volume}{193}, \bibinfo{pages}{105844}.
\bibitem[{Mandriota et~al.(2024a)Mandriota, Menga and Carbone}]{mandriota2024adhesive}
\bibinfo{author}{Mandriota, C.}, \bibinfo{author}{Menga, N.}, \bibinfo{author}{Carbone, G.}, \bibinfo{year}{2024}a.
\newblock \bibinfo{title}{Adhesive contact mechanics of viscoelastic materials}.
\newblock \bibinfo{journal}{International Journal of Solids and Structures} \bibinfo{volume}{290}, \bibinfo{pages}{112685}.
\bibitem[{Mandriota et~al.(2024b)Mandriota, Menga and Carbone}]{mandriota2024enhancement}
\bibinfo{author}{Mandriota, C.}, \bibinfo{author}{Menga, N.}, \bibinfo{author}{Carbone, G.}, \bibinfo{year}{2024}b.
\newblock \bibinfo{title}{Enhancement of adhesion strength in viscoelastic unsteady contacts}.
\newblock \bibinfo{journal}{Journal of the Mechanics and Physics of Solids} \bibinfo{volume}{192}, \bibinfo{pages}{105826}.
\bibitem[{Maugis(2013)}]{maugis2013contact}
\bibinfo{author}{Maugis, D.}, \bibinfo{year}{2013}.
\newblock \bibinfo{title}{Contact, adhesion and rupture of elastic solids}. volume \bibinfo{volume}{130}.
\newblock \bibinfo{publisher}{Springer Science \& Business Media}.
\bibitem[{Mazzolai et~al.(2019)Mazzolai, Mondini, Tramacere, Riccomi, Sadeghi, Giordano, Del~Dottore, Scaccia, Zampato and Carminati}]{mazzolai2019octopus}
\bibinfo{author}{Mazzolai, B.}, \bibinfo{author}{Mondini, A.}, \bibinfo{author}{Tramacere, F.}, \bibinfo{author}{Riccomi, G.}, \bibinfo{author}{Sadeghi, A.}, \bibinfo{author}{Giordano, G.}, \bibinfo{author}{Del~Dottore, E.}, \bibinfo{author}{Scaccia, M.}, \bibinfo{author}{Zampato, M.}, \bibinfo{author}{Carminati, S.}, \bibinfo{year}{2019}.
\newblock \bibinfo{title}{Octopus-inspired soft arm with suction cups for enhanced grasping tasks in confined environments}.
\newblock \bibinfo{journal}{Advanced Intelligent Systems} \bibinfo{volume}{1}, \bibinfo{pages}{1900041}.
\bibitem[{M{\"u}ser and Persson(2022)}]{muser2022crack}
\bibinfo{author}{M{\"u}ser, M.H.}, \bibinfo{author}{Persson, B.N.}, \bibinfo{year}{2022}.
\newblock \bibinfo{title}{Crack and pull-off dynamics of adhesive, viscoelastic solids}.
\newblock \bibinfo{journal}{Europhysics Letters} \bibinfo{volume}{137}, \bibinfo{pages}{36004}.
\bibitem[{Narkar et~al.(2019)Narkar, Kendrick, Bellur, Leftwich, Zhang and Lee}]{Narkar2019}
\bibinfo{author}{Narkar, A.R.}, \bibinfo{author}{Kendrick, C.}, \bibinfo{author}{Bellur, K.}, \bibinfo{author}{Leftwich, T.}, \bibinfo{author}{Zhang, Z.}, \bibinfo{author}{Lee, B.P.}, \bibinfo{year}{2019}.
\newblock \bibinfo{title}{Rapidly responsive smart adhesive-coated micropillars utilizing catechol--boronate complexation chemistry}.
\newblock \bibinfo{journal}{Soft matter} \bibinfo{volume}{15}, \bibinfo{pages}{5474--5482}.
\bibitem[{Nazari et~al.(2024)Nazari, Papangelo and Ciavarella}]{nazari2024friction}
\bibinfo{author}{Nazari, R.}, \bibinfo{author}{Papangelo, A.}, \bibinfo{author}{Ciavarella, M.}, \bibinfo{year}{2024}.
\newblock \bibinfo{title}{Friction in rolling a cylinder on or under a viscoelastic substrate with adhesion}.
\newblock \bibinfo{journal}{Tribology Letters} \bibinfo{volume}{72}, \bibinfo{pages}{50}.
\bibitem[{Oliver et~al.(2023)Oliver, Dalmas and Scheibert}]{oliver2023adhesion}
\bibinfo{author}{Oliver, C.}, \bibinfo{author}{Dalmas, D.}, \bibinfo{author}{Scheibert, J.}, \bibinfo{year}{2023}.
\newblock \bibinfo{title}{Adhesion in soft contacts is minimum beyond a critical shear displacement}.
\newblock \bibinfo{journal}{Journal of the Mechanics and Physics of Solids} \bibinfo{volume}{181}, \bibinfo{pages}{105445}.
\bibitem[{Papangelo and Ciavarella(2023)}]{papangelo2023detachment}
\bibinfo{author}{Papangelo, A.}, \bibinfo{author}{Ciavarella, M.}, \bibinfo{year}{2023}.
\newblock \bibinfo{title}{Detachment of a rigid flat punch from a viscoelastic material}.
\newblock \bibinfo{journal}{Tribology letters} \bibinfo{volume}{71}, \bibinfo{pages}{48}.
\bibitem[{Papangelo et~al.(2024)Papangelo, Nazari and Ciavarella}]{Papangelo2024friction}
\bibinfo{author}{Papangelo, A.}, \bibinfo{author}{Nazari, R.}, \bibinfo{author}{Ciavarella, M.}, \bibinfo{year}{2024}.
\newblock \bibinfo{title}{Friction for a sliding adhesive viscoelastic cylinder: Effect of maugis parameter}.
\newblock \bibinfo{journal}{European Journal of Mechanics - A/Solids} \bibinfo{volume}{107}, \bibinfo{pages}{105348}.
\newblock \URLprefix \url{https://www.sciencedirect.com/science/article/pii/S0997753824001281}, \DOIprefix\doi{https://doi.org/10.1016/j.euromechsol.2024.105348}.
\bibitem[{Papangelo et~al.(2019)Papangelo, Scheibert, Sahli, Pallares and Ciavarella}]{papangelo2019shear}
\bibinfo{author}{Papangelo, A.}, \bibinfo{author}{Scheibert, J.}, \bibinfo{author}{Sahli, R.}, \bibinfo{author}{Pallares, G.}, \bibinfo{author}{Ciavarella, M.}, \bibinfo{year}{2019}.
\newblock \bibinfo{title}{Shear-induced contact area anisotropy explained by a fracture mechanics model}.
\newblock \bibinfo{journal}{Physical Review E} \bibinfo{volume}{99}, \bibinfo{pages}{053005}.
\bibitem[{Perriot and Barthel(2004)}]{perriot2004elastic}
\bibinfo{author}{Perriot, A.}, \bibinfo{author}{Barthel, E.}, \bibinfo{year}{2004}.
\newblock \bibinfo{title}{Elastic contact to a coated half-space: Effective elastic modulus and real penetration}.
\newblock \bibinfo{journal}{Journal of Materials Research} \bibinfo{volume}{19}, \bibinfo{pages}{600--608}.
\bibitem[{Persson and Brener(2005)}]{persson2005crack}
\bibinfo{author}{Persson, B.}, \bibinfo{author}{Brener, E.}, \bibinfo{year}{2005}.
\newblock \bibinfo{title}{Crack propagation in viscoelastic solids}.
\newblock \bibinfo{journal}{Physical Review E} \bibinfo{volume}{71}, \bibinfo{pages}{036123}.
\bibitem[{Persson(2024)}]{persson2024influencetemperaturecracktipspeed}
\bibinfo{author}{Persson, B.N.J.}, \bibinfo{year}{2024}.
\newblock \bibinfo{title}{Influence of temperature and crack-tip speed on crack propagation in elastic solids}.
\newblock \URLprefix \url{https://arxiv.org/abs/2409.06182}, \href{http://arxiv.org/abs/2409.06182}{{\tt arXiv:2409.06182}}.
\bibitem[{Qin et~al.(2024)Qin, Zhang, Tan, Yang, Wang, Zhang, Wang and Liu}]{Qin2024}
\bibinfo{author}{Qin, H.}, \bibinfo{author}{Zhang, C.}, \bibinfo{author}{Tan, W.}, \bibinfo{author}{Yang, L.}, \bibinfo{author}{Wang, R.}, \bibinfo{author}{Zhang, Y.}, \bibinfo{author}{Wang, F.}, \bibinfo{author}{Liu, L.}, \bibinfo{year}{2024}.
\newblock \bibinfo{title}{Bionic adhesion systems: From natural design to artificial application}.
\newblock \bibinfo{journal}{Advanced Materials Technologies} \bibinfo{volume}{9}, \bibinfo{pages}{2301387}.
\bibitem[{Qu et~al.(2024)Qu, Yu, Tang, Xu, Mao and Zhou}]{qu2024advanced}
\bibinfo{author}{Qu, J.}, \bibinfo{author}{Yu, Z.}, \bibinfo{author}{Tang, W.}, \bibinfo{author}{Xu, Y.}, \bibinfo{author}{Mao, B.}, \bibinfo{author}{Zhou, K.}, \bibinfo{year}{2024}.
\newblock \bibinfo{title}{Advanced technologies and applications of robotic soft grippers}.
\newblock \bibinfo{journal}{Advanced Materials Technologies} \bibinfo{volume}{9}, \bibinfo{pages}{2301004}.
\bibitem[{Sahli et~al.(2019)Sahli, Pallares, Papangelo, Ciavarella, Ducottet, Ponthus and Scheibert}]{sahli2019shear}
\bibinfo{author}{Sahli, R.}, \bibinfo{author}{Pallares, G.}, \bibinfo{author}{Papangelo, A.}, \bibinfo{author}{Ciavarella, M.}, \bibinfo{author}{Ducottet, C.}, \bibinfo{author}{Ponthus, N.}, \bibinfo{author}{Scheibert, J.}, \bibinfo{year}{2019}.
\newblock \bibinfo{title}{Shear-induced anisotropy in rough elastomer contact}.
\newblock \bibinfo{journal}{Physical Review Letters} \bibinfo{volume}{122}, \bibinfo{pages}{214301}.
\bibitem[{Schapery(1975a)}]{schapery1975theory1}
\bibinfo{author}{Schapery, R.A.}, \bibinfo{year}{1975}a.
\newblock \bibinfo{title}{A theory of crack initiation and growth in viscoelastic media: I. theoretical development}.
\newblock \bibinfo{journal}{International Journal of fracture} \bibinfo{volume}{11}, \bibinfo{pages}{141--159}.
\bibitem[{Schapery(1975b)}]{schapery1975theory2}
\bibinfo{author}{Schapery, R.A.}, \bibinfo{year}{1975}b.
\newblock \bibinfo{title}{A theory of crack initiation and growth in viscoelastic media ii. approximate methods of analysis}.
\newblock \bibinfo{journal}{International Journal of Fracture} \bibinfo{volume}{11}, \bibinfo{pages}{369--388}.
\bibitem[{Schapery(2022)}]{schapery2022theory}
\bibinfo{author}{Schapery, R.A.}, \bibinfo{year}{2022}.
\newblock \bibinfo{title}{A theory of viscoelastic crack growth: revisited}.
\newblock \bibinfo{journal}{International Journal of Fracture} \bibinfo{volume}{233}, \bibinfo{pages}{1--16}.
\bibitem[{Shintake et~al.(2018)Shintake, Cacucciolo, Floreano and Shea}]{shintake2018soft}
\bibinfo{author}{Shintake, J.}, \bibinfo{author}{Cacucciolo, V.}, \bibinfo{author}{Floreano, D.}, \bibinfo{author}{Shea, H.}, \bibinfo{year}{2018}.
\newblock \bibinfo{title}{Soft robotic grippers}.
\newblock \bibinfo{journal}{Advanced materials} \bibinfo{volume}{30}, \bibinfo{pages}{1707035}.
\bibitem[{Shui et~al.(2020)Shui, Jia, Li, Guo, Guo, Liu, Liu and Chen}]{Shui2020}
\bibinfo{author}{Shui, L.}, \bibinfo{author}{Jia, L.}, \bibinfo{author}{Li, H.}, \bibinfo{author}{Guo, J.}, \bibinfo{author}{Guo, Z.}, \bibinfo{author}{Liu, Y.}, \bibinfo{author}{Liu, Z.}, \bibinfo{author}{Chen, X.}, \bibinfo{year}{2020}.
\newblock \bibinfo{title}{Rapid and continuous regulating adhesion strength by mechanical micro-vibration}.
\newblock \bibinfo{journal}{Nature communications} \bibinfo{volume}{11}, \bibinfo{pages}{1583}.
\bibitem[{Tabor(1977)}]{tabor1977surface}
\bibinfo{author}{Tabor, D.}, \bibinfo{year}{1977}.
\newblock \bibinfo{title}{Surface forces and surface interactions}.
\newblock \bibinfo{journal}{Journal of Colloid and Interface Science} \bibinfo{volume}{58}, \bibinfo{pages}{2--13}.
\bibitem[{Tiwari et~al.(2017)Tiwari, Dorogin, Bennett, Schulze, Sawyer, Tahir, Heinrich and Persson}]{tiwari2017effect}
\bibinfo{author}{Tiwari, A.}, \bibinfo{author}{Dorogin, L.}, \bibinfo{author}{Bennett, A.}, \bibinfo{author}{Schulze, K.}, \bibinfo{author}{Sawyer, W.}, \bibinfo{author}{Tahir, M.}, \bibinfo{author}{Heinrich, G.}, \bibinfo{author}{Persson, B.}, \bibinfo{year}{2017}.
\newblock \bibinfo{title}{The effect of surface roughness and viscoelasticity on rubber adhesion}.
\newblock \bibinfo{journal}{Soft matter} \bibinfo{volume}{13}, \bibinfo{pages}{3602--3621}.
\bibitem[{Trivedi et~al.(2008)Trivedi, Rahn, Kier and Walker}]{trivedi2008soft}
\bibinfo{author}{Trivedi, D.}, \bibinfo{author}{Rahn, C.D.}, \bibinfo{author}{Kier, W.M.}, \bibinfo{author}{Walker, I.D.}, \bibinfo{year}{2008}.
\newblock \bibinfo{title}{Soft robotics: Biological inspiration, state of the art, and future research}.
\newblock \bibinfo{journal}{Applied bionics and biomechanics} \bibinfo{volume}{5}, \bibinfo{pages}{99--117}.
\bibitem[{VanDonselaar et~al.(2023)VanDonselaar, Bellido-Aguilar, Safaripour, Kim, Watkins, Crosby, Webster and Croll}]{vandonselaar2023silicone}
\bibinfo{author}{VanDonselaar, K.R.}, \bibinfo{author}{Bellido-Aguilar, D.A.}, \bibinfo{author}{Safaripour, M.}, \bibinfo{author}{Kim, H.}, \bibinfo{author}{Watkins, J.J.}, \bibinfo{author}{Crosby, A.J.}, \bibinfo{author}{Webster, D.C.}, \bibinfo{author}{Croll, A.B.}, \bibinfo{year}{2023}.
\newblock \bibinfo{title}{Silicone elastomers and the persson-brener adhesion model}.
\newblock \bibinfo{journal}{The Journal of Chemical Physics} \bibinfo{volume}{159}.
\bibitem[{Violano and Afferrante(2022)}]{violano2022size}
\bibinfo{author}{Violano, G.}, \bibinfo{author}{Afferrante, L.}, \bibinfo{year}{2022}.
\newblock \bibinfo{title}{Size effects in adhesive contacts of viscoelastic media}.
\newblock \bibinfo{journal}{European Journal of Mechanics-A/Solids} \bibinfo{volume}{96}, \bibinfo{pages}{104665}.
\bibitem[{Violano et~al.(2021)Violano, Chateauminois and Afferrante}]{Violano2021rate}
\bibinfo{author}{Violano, G.}, \bibinfo{author}{Chateauminois, A.}, \bibinfo{author}{Afferrante, L.}, \bibinfo{year}{2021}.
\newblock \bibinfo{title}{Rate-dependent adhesion of viscoelastic contacts, part i: Contact area and contact line velocity within model randomly rough surfaces}.
\newblock \bibinfo{journal}{Mechanics of Materials} \bibinfo{volume}{160}, \bibinfo{pages}{103926}.
\newblock \URLprefix \url{https://www.sciencedirect.com/science/article/pii/S0167663621001708}, \DOIprefix\doi{https://doi.org/10.1016/j.mechmat.2021.103926}.
\bibitem[{Wahl et~al.(2006)Wahl, Asif, Greenwood and Johnson}]{wahl2006oscillating}
\bibinfo{author}{Wahl, K.}, \bibinfo{author}{Asif, S.}, \bibinfo{author}{Greenwood, J.}, \bibinfo{author}{Johnson, K.}, \bibinfo{year}{2006}.
\newblock \bibinfo{title}{Oscillating adhesive contacts between micron-scale tips and compliant polymers}.
\newblock \bibinfo{journal}{Journal of colloid and interface science} \bibinfo{volume}{296}, \bibinfo{pages}{178--188}.
\bibitem[{Wang et~al.(2019)Wang, Kang, Arzt, Federle and Hensel}]{wang2019strong}
\bibinfo{author}{Wang, Y.}, \bibinfo{author}{Kang, V.}, \bibinfo{author}{Arzt, E.}, \bibinfo{author}{Federle, W.}, \bibinfo{author}{Hensel, R.}, \bibinfo{year}{2019}.
\newblock \bibinfo{title}{Strong wet and dry adhesion by cupped microstructures}.
\newblock \bibinfo{journal}{ACS applied materials \& interfaces} \bibinfo{volume}{11}, \bibinfo{pages}{26483--26490}.
\bibitem[{Williams et~al.(1955)Williams, Landel and Ferry}]{WLF}
\bibinfo{author}{Williams, M.L.}, \bibinfo{author}{Landel, R.F.}, \bibinfo{author}{Ferry, J.D.}, \bibinfo{year}{1955}.
\newblock \bibinfo{title}{The temperature dependence of relaxation mechanisms in amorphous polymers and other glass-forming liquids}.
\newblock \bibinfo{journal}{Journal of the American Chemical society} \bibinfo{volume}{77}, \bibinfo{pages}{3701--3707}.
\bibitem[{Yi et~al.(2024)Yi, Haouas, Gauthier and Rabenorosoa}]{Yi2024}
\bibinfo{author}{Yi, J.}, \bibinfo{author}{Haouas, W.}, \bibinfo{author}{Gauthier, M.}, \bibinfo{author}{Rabenorosoa, K.}, \bibinfo{year}{2024}.
\newblock \bibinfo{title}{A pdms/silicon adhesion control method at millimeter-scale based on microvibration}.
\newblock \bibinfo{journal}{Advanced Intelligent Systems} , \bibinfo{pages}{2400394}.
\bibitem[{Zhao et~al.(2022)Zhao, Li, Tan, Liu, Lu and Shi}]{Zhao2022}
\bibinfo{author}{Zhao, J.}, \bibinfo{author}{Li, X.}, \bibinfo{author}{Tan, Y.}, \bibinfo{author}{Liu, X.}, \bibinfo{author}{Lu, T.}, \bibinfo{author}{Shi, M.}, \bibinfo{year}{2022}.
\newblock \bibinfo{title}{Smart adhesives via magnetic actuation}.
\newblock \bibinfo{journal}{Advanced Materials} \bibinfo{volume}{34}, \bibinfo{pages}{2107748}.

\end{thebibliography}
\end{document}